# FieldRT: Reconfigurable Architecture from Geometric Sensitivity to Streaming Acceleration

Haichuan Li
haichuan.li@utu.fi

## Abstract

Geometric perception on edge must simultaneously achieve two goals that are rarely addressed together: preserve accurate spatial understanding for object localization and operate within stringent latency, memory, energy, and precision constraints. This coupling is particularly evident in oriented bird's-eye-view (BEV) detection, where global context, multi-scale interactions, and geometric fusion affect both prediction quality and hardware dataflow. Conventional deployment flows first compress a model using aggregate quality or complexity proxies and then map the compressed operations to a fixed accelerator interface. This separation risks removing geometric structures that appear redundant under coarse metrics but are critical for precise localization, while leaving hardware-level efficiencies in the retained geometry unexploited. We present FieldRT, an algorithm-hardware co-design methodology that uses a common analysis of geometric sensitivity and data lifetime to derive both the deployment profile and its spatial realization. We deploy FieldRT with Axis-Fusion Network (AFN), a sub-million-parameter oriented-BEV detector, and construct controlled profiles that vary the capacity and organization of its geometric processing. The experiments show that redundancy is task-dependent rather than uniform across metrics. Dense prediction can improve while real instance detection deteriorates; coarse localization may remain stable while strict-IoU precision degrades. Moreover, transformations that are individually safe can become detrimental when combined. FieldRT translates the retained geometric structure into a parameterized streaming architecture that shares field access and storage across spatial operators. The resulting custom-PL primitives achieve production-scale numerical closure on an AMD Kria KV260 and make otherwise infeasible multi-dilation execution realizable. Complete TensorRT deployments on an NVIDIA Jetson AGX Orin validate the selected profiles across simulation and real workloads. They also reveal numerical failures that are hidden when only checking whether the model compiles and runs succesfully. FieldRT demonstrates that geometry-preserving profile construction and efficient spatial realization must be derived jointly from prediction sensitivity, numerical behavior, producer–consumer structure, and data lifetime.

## 1 Introduction

Geometric perception on edge devices must preserve object position, extent, class, and orientation under constraints on latency, memory, energy, and numerical precision. These requirements link prediction semantics with execution structure. The model components that preserve strict instance geometry determine which spatial information must remain available, while hardware decisions, such as numerical representation, buffering, data transfer, and scheduling, determine whether that information remains intact through deployment.

Conventional deployment workflows separate these decisions. A model is first compressed using aggregate proxies such as parameter count, FLOPs, dense prediction quality, or isolated accelerator latency. The resulting operator graph is then mapped to a predefined hardware interface. Model-level proxies do not reveal which structures preserve instance geometry, whether their redundancy depends on the surrounding configuration, or whether a selected profile remains reliable across workloads. Operator-level mappings likewise obscure shared field access, producer–consumer locality, and spatial data lifetime. Consequently, a profile that appears efficient under aggregate metrics can lose strict geometric quality, while a graph that maps poorly to a fixed accelerator can underutilize its resources, e.g., computation, memory, power, as compared to a programmable accelerator.

We study this problem using Axis-Fusion Network (AFN), a compact oriented bird's-eye-view detector that converts point-cloud observations into fixed-resolution geometric fields. AFN combines complementary axis-conditioned views, adaptive geometric fusion, multi-scale spatial processing, global context, and oriented prediction. Although AFN contains fewer than one million parameters and operates on statically shaped tensors, its deployment behavior depends on the geometric roles and interactions of its components rather than on model size alone.

Controlled transformations of AFN reveal that deployment-relevant redundancy is conditional rather than uniform. Dense prediction, coarse detection, strict localization, orientation estimation, and cross-workload robustness respond differently to changes in context, fusion, spatial scale, and model capacity. Transformations that are effective independently can also interact negatively when combined. A reliable deployment profile must therefore be selected from direct evidence of instance-level geometric preservation rather than from uniform compression or aggregate prediction metrics. This geometric structure also determines how the retained computation can be realized efficiently.

Axis-conditioned branches consume common field data, multi-dilation neighborhoods reuse overlapping spatial state, and global context introduces reductions that can be accumulated during streaming. These relationships expose opportunities for single-read distribution, streaming fusion, shared buffering, and persistent aggregation. A conventional operator interface does not capture these opportunities and can fragment a compact, statically bounded graph across accelerator and host execution, interrupting locality and materializing intermediate tensors.

We present FieldRT, an algorithm-hardware co-design methodology that jointly derives a geometry-preserving deployment profile and its spatial architecture. FieldRT first measures how controlled structural transformations affect dense recognition, oriented instances, strict localization, and cross-workload behavior.

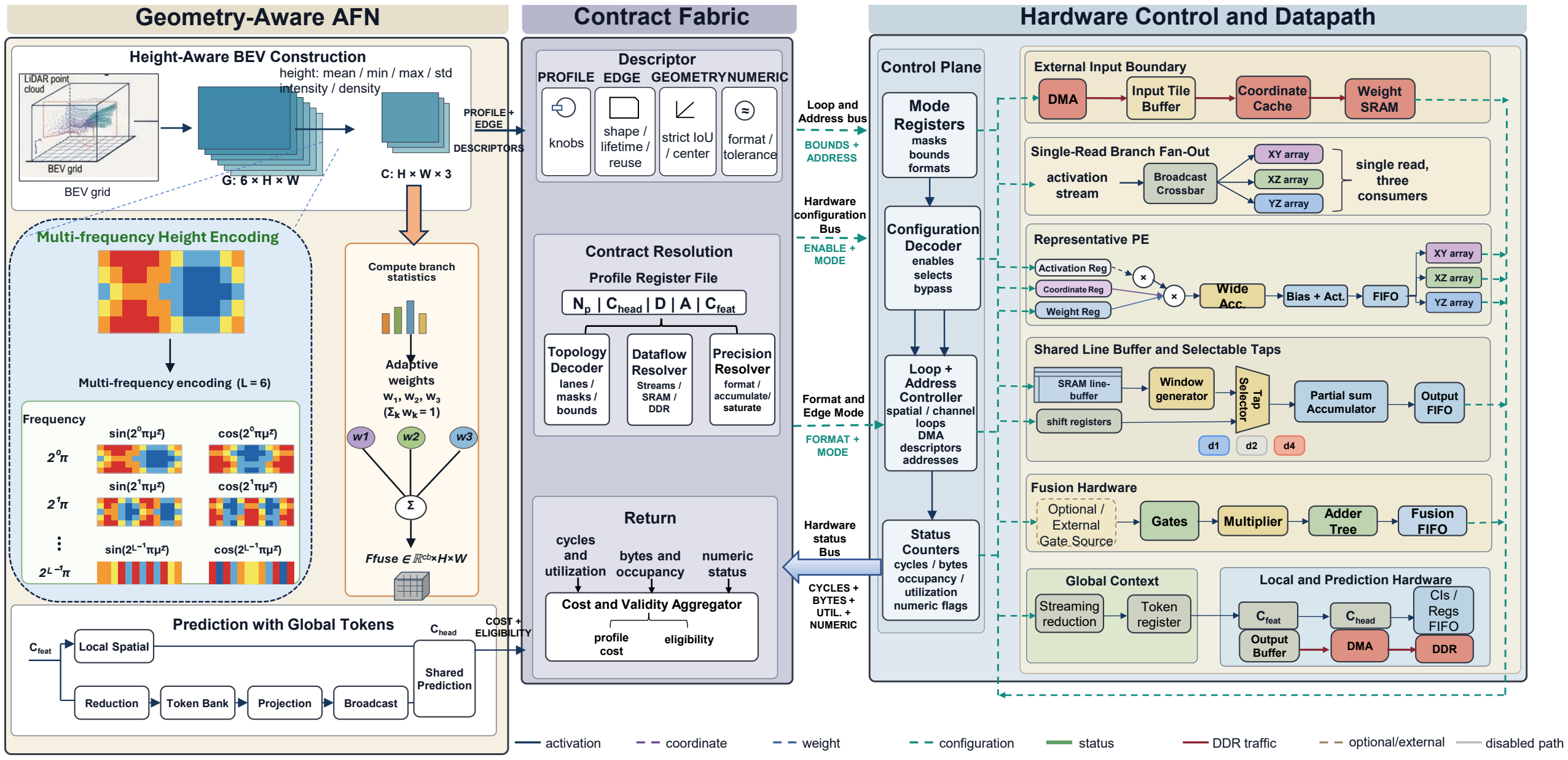

**Figure 1: FieldRT algorithm-hardware co-design framework. Controlled AFN profiles expose geometry-preserving structure through profile and edge descriptors. The contract fabric translates these descriptors into hardware topology, numerical format, buffering, control, and streaming dataflow, and returns measured execution cost and validity to the profile selection process.**

It then represents the retained computation through a geometric execution contract that captures field semantics, numerical format, producer–consumer relationships, traversal patterns, and state lifetime. The same contract guides both profile selection and hardware organization. FieldRT lowers common geometric structures into parameterized spatial execution. Shared fields become single-read multi-consumer streams, axis-conditioned responses become streaming fusion, retained dilation patterns determine line-buffer organization, and global aggregation becomes streaming state. Profile choices therefore configure branch topology, numerical representation, buffer lifetime, communication boundaries, and parallelism together, rather than modifying the model independently of its execution architecture. We instantiate FieldRT through two complementary deployment paths. On an AMD Kria KV260, a geometry-native custom-PL architecture realizes single-read field distribution, streaming fusion, lifetime-aware multi-dilation buffering, and streaming reduction. On an NVIDIA Jetson AGX Orin, complete TensorRT deployments evaluate the selected profiles under full task, numerical, latency, power, and energy boundaries. The programmable-logic experiments establish the resource and dataflow consequences of the retained geometry, while the complete-model deployments establish how profile choices affect instance quality and system efficiency across Simulation and Real workloads. The combined results show that geometric deployment cannot be optimized using a scalar measure of model complexity or an operator graph alone. The relevant design object is the relationship between the structures that preserve instance geometry and the numerical state, storage lifetime, communication, and producer–consumer dataflow through which those structures are executed. FieldRT makes this relationship explicit and uses it to coordinate model profiling, spatial architecture, and deployment validation. Our work makes the following contributions:

- **Conditional geometric redundancy.** We show that dense prediction, instance detection, strict localization, orientation, and cross-workload robustness respond differently to structural transformations, and that independently effective transformations can interact negatively.
- **FieldRT algorithm-hardware co-design.** We introduce a unified design that jointly derives a geometry-preserving deployment profile and its spatial execution architecture from geometric sensitivity, producer–consumer structure, numerical representation, and data lifetime.
- **Geometry-sensitive profile.** We develop a controlled profiling procedure that selects deployment configurations using instance-level geometry and transformation interactions rather than uniform complexity or aggregate-quality proxies.
- **Geometry-native spatial architecture.** We translate shared fields, axis-conditioned processing, multi-dilations, and global aggregation into single-read distribution, streaming fusion, lifetime-aware buffering, and streaming reduction.
- **Cross-platform realization and evaluation.** We evaluate complete TensorRT deployments on an NVIDIA Jetson AGX Orin and production-scale programmable-logic execution on an AMD Kria KV260 across Simulation and Real workloads.

## 2 Background and related works

*Efficient LiDAR and BEV Perception.* Point-based methods, including PointNet [22], PointNet++ [23], DGCNN [30], KPConv [27],

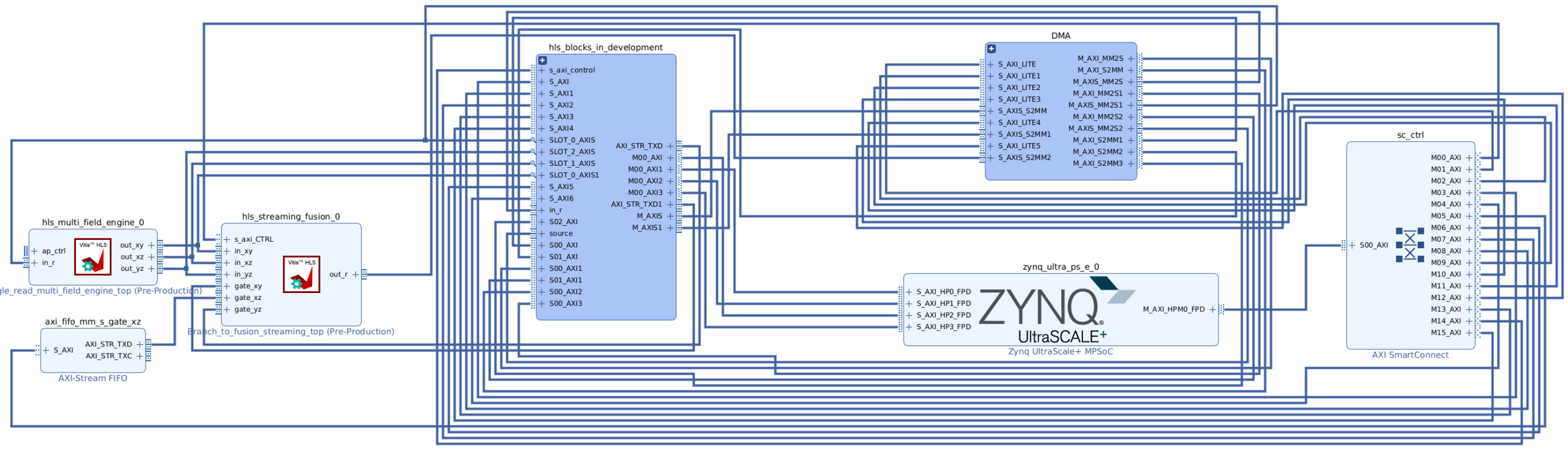

**Figure 2: Integrated FieldRT platform on the AMD Kria KV260. The Vivado block design connects the single-read Multi-Field Engine to the gated streaming-fusion path and integrates the Zynq UltraScale+ processing system, AXI control interconnect, multiple MM2S/S2MM DMA channels, and the extended HLS kernel fabric. The implemented platform supports the production-scale programmable-logic experiments evaluated in this section.**

RandLA-Net [12], and Point Transformer [35], learn directly from irregular point sets and preserve detailed geometric information. Voxel-based and sparse-convolutional methods, such as VoxelNet [37], SECOND [31], SparseConvNet [10], MinkowskiNet [7], and Cylinder3D [38], instead discretize 3D space to support structured feature extraction. Although effective, point-neighborhood aggregation and 3D feature processing can introduce substantial computational and memory overhead. BEV and pillar-based representations reduce this cost by transforming point clouds into structured 2D feature maps. Representative approaches include MV3D [6], PIXOR [32], PointPillars [13], CenterPoint [34], PillarNet [24], and BEV-MAE [5]. Related BEV methods have also explored temporal reasoning, depth-aware projection, multi-modal fusion, and dense scene representations [4, 16, 17, 19, 26, 36].

FPGA deployment frameworks exposes a fundamental trade-off between rapid compilation and operator coverage. Vitis AI maps supported operators efficiently to the Deep-learning Processing Unit (DPU), but unsupported operators execute on the CPU, introducing data-transfer, synchronization, and scheduling overheads. Consequently, Zynq- and Kria-class platforms achieve strong acceleration for DPU-resident layers but can degrade substantially when execution crosses the CPU–DPU boundary. Existing studies primarily report end-to-end latency or throughput, while the default Vitis AI runtime provides limited attribution of time to individual DPU and CPU subgraphs. FieldRT therefore profiles the system-level cost of fallback for a non-conforming BEV model on the KV260, rather than redesigning the network solely to maximize DPU residency. Quantization alone does not resolve this compatibility problem. FPGA deployments of 3D detectors show that sparse, scatter, voxel, and geometry-processing operators may still require software execution or custom RTL after INT8 quantization. LiFT demonstrates the complementary approach of co-designing the model and inference pipeline around DPU and FINN constraints to avoid fallback. In contrast, FieldRT preserves an accuracy-first BEV architecture with minimal modification, exposing the deployment-fidelity cost of the commercial compatibility path. The KV260 is thus an experimental vehicle, not an absolute performance baseline. Its fixed DPU abstraction and reconfigurable fabric permit compiler-directed heterogeneous execution and custom hardware realization to be studied on the same silicon, helping separate software-stack effects from underlying FPGA capability. The AGX Orin instead provides a production-grade TensorRT reference for end-to-end deployment. Together, these platforms reveal which AFN structures map through default flows, when fallback becomes the dominant bottleneck, and how selected geometry-native operations can be reformulated as streaming FPGA kernels with explicit numerical and execution evidence. The comparison targets deployment fidelity, observability, and hardware–software mapping strategy rather than raw accelerator performance [1–3, 8, 9, 14, 15, 18, 20, 21, 25, 28, 29, 33].

## 3 Algorithm Methodology

Our FieldRT couples two decisions that conventional deployment workflows address separately: identifying the model structures required to preserve strict instance geometry and deriving an execution architecture for those structures. Figure 1 illustrates this process. It first represents geometric computation through spatial fields, constructs controlled profiles from a common reference model, and measures the geometric effects of both individual and combined transformations. The retained profile is then lowered to spatial hardware through an execution contract that records details.

### 3.1 Field-Based BEV Perception

For frame $t$, a deterministic preprocessing operator maps an irregular point set $\mathcal{P}_t$ to a fixed BEV tensor

$$\mathbf{X}_t = \mathcal{V}(\mathcal{P}_t) \in \mathbb{R}^{6\times H\times W}, \qquad H = W = 400.$$

The six channels contain cell-wise mean, minimum, maximum, and standard deviation of height, mean intensity, and normalized point density. An aligned coordinate tensor

$$\mathbf{C}_t = [\mathbf{C}^x, \mathbf{C}^y, \mathbf{C}^z_t] \in \mathbb{R}^{H\times W\times 3}$$

contains static planar coordinates and a frame-dependent mean-height field. The designed AFN predicts

$$(\mathbf{Y}_t^{\mathrm{cls}}, \mathbf{Y}_t^{\mathrm{reg}}) = f_\theta(\mathbf{X}_t, \mathbf{C}_t),$$

where classification has $K_d + 1$ workload-dependent channels and regression encodes $(\Delta x, \Delta y, \log l, \log w, \cos 2\theta, \sin 2\theta)$ under a $\pi$-periodic orientation convention.

We use *geometric field* to denote a spatially aligned tensor with explicit geometric semantics and characterize it by its spatial meaning, static or frame dependence, consumer set, and producer–consumer lifetime. These properties expose reuse and buffering opportunities that tensor shape alone does not represent.

## 3.2 Axis-Fusion Network

AFN augments the six-channel BEV tensor with a weighted multi-frequency encoding of the frame-dependent height field:

$$\mathbf{F}_t^0 = \mathrm{Concat}\left[\mathbf{X}_t,\ \mathbf{w} \odot \{\sin(2^\ell \pi \mathbf{C}_t^z), \cos(2^\ell \pi \mathbf{C}_t^z)\}_{\ell=0}^{L-1}\right], L = 6, \quad (1)$$

where $w_j = \exp(-j/2)$. Three independently parameterized branches process the augmented field under coordinate modulation:

$$\mathbf{F}_t^b = \phi_b(\mathbf{F}_t^0) \odot [1 + \Gamma_b(\mathbf{C}_t; \alpha_b)], \qquad b \in \{\mathrm{xy}, \mathrm{xz}, \mathrm{yz}\}, \quad (2)$$

where $\phi_b$ is Conv–BN–ReLU and $\Gamma_b$ combines the height field with the corresponding planar coordinate. Appendix C gives the exact modulation functions. A denoising module shared across the three perspective branches applies parallel dilated convolutions:

$$\mathcal{M}_{\mathcal{D}_p}(\mathbf{F}) = \mathbf{F} + \mathrm{Conv}_{1\times1}^p \left( \Big\Vert_{d \in \mathcal{D}_p} \mathcal{D}_d^p(\mathbf{F}) \right). \quad (3)$$

AFN combines the denoised branches using global variance-derived weights and a pairwise consistency coefficient:

$$\mathbf{F}_t^{\mathrm{enc}} = \kappa_t\, \mathrm{BN}\left( \sum_b \frac{\exp(-v_b) a_b}{\sum_j \exp(-v_j) a_j + \epsilon} \widehat{\mathbf{F}}_t^b \right), \frac{\exp(-v_b)\, a_b}{\max\left( \sum_j \exp(-v_j)\, a_j,\ \epsilon \right)}, v_b = \mathrm{Var}(\mathbf{F}_t^b). \quad (4)$$

A lightweight encoder–decoder aggregates local multiresolution context, while pooled encoder features and learned embeddings form $N_p$ global tokens. Their broadcast representation is fused with the local feature and processed by separate classification and double-angle regression heads. Complete architectural and training details appear in Appendix C.

## 3.3 Geometry-Sensitive Profile Construction

We represent an AFN profile as:

$$p = (N_p, C_p^{\mathrm{head}}, \mathcal{D}_p, \mathcal{A}_p, C_p^{\mathrm{feat}}), \quad (5)$$

covering global-token count, prediction-head width, active dilations, geometric-axis fusion, and common feature width. Starting from Quality, we evaluate isolated and combined token, head, dilation, fusion, and feature-width transformations. D14 removes the intermediate $d = 2$ branch while retaining $d \in \{1, 4\}$; Feat-48 and the fusion variants are used for sensitivity analysis but are not included in complete TensorRT deployment.

For profile $p$ and workload $d$, FieldRT retains:

$$\mathbf{q}_{p,d} = [\mathrm{mAP}, \mathrm{AP}_{50}, \mathrm{AP}_{75}, \mathrm{mIoU}, \mathrm{Dense\text{-}mAP}, e_{\mathrm{yaw}}], \quad (6)$$

rather than reducing geometric quality to a scalar. This distinction captures disagreements between dense recognition, permissive overlap, strict localization, and periodic orientation error. For directly evaluated transformations $a$ and $b$, their interaction is

$$I_j(a, b; r, d) = \Delta_j(a \oplus b; r, d) - \Delta_j(a; r, d) - \Delta_j(b; r, d), \quad (7)$$

where $\Delta_j$ is the signed normalized change from the matched Quality reference, with positive values denoting improvement. Negative $I_j$ indicates that the combined profile underperforms the additive expectation. Interactions are computed within each workload under matched protocols and are descriptive rather than predictive.

**Table 1: AFN profiles used for TensorRT deployment.**

| Profile | Tokens | Head | Dilations | Fusion | Width |
|---|---|---|---|---|---|
| Quality | 4 | 32 | $\{1, 2, 4\}$ | Geo. | 64 |
| Tokens-2 | 2 | 32 | $\{1, 2, 4\}$ | Geo. | 64 |
| Head-16 | 4 | 16 | $\{1, 2, 4\}$ | Geo. | 64 |
| D14 | 4 | 32 | $\{1, 4\}$ | Geo. | 64 |
| H16-D14 | 4 | 16 | $\{1, 4\}$ | Geo. | 64 |

## 3.4 Geometric Execution Contract

For each retained profile $p$, FieldRT constructs a geometric execution contract

$$C_p = (\mathcal{F}_p, \mathcal{E}_p, \mathcal{N}_p, \mathcal{T}_p, \mathcal{L}_p), \quad (8)$$

where $\mathcal{F}_p$ describes field semantics and spatial extent, $\mathcal{E}_p$ captures producer–consumer relationships and reuse, $\mathcal{N}_p$ specifies numerical representation and accumulation requirements, $\mathcal{T}_p$ defines traversal and branch topology, and $\mathcal{L}_p$ describes state lifetime. The contract provides the design interface between geometry-sensitive profile construction and backend-specific spatial realization. FieldRT resolves the contract into four spatial execution patterns. Shared geometric fields induce single-read multi-consumer streams, axis-conditioned branches induce gated streaming fusion, multi-dilation neighborhoods induce lifetime-shared line storage, and global aggregation becomes persistent streaming state. For profile $p$, the largest active dilation and retained vertical span are

$$D_{\max}(p) = \max_{d \in \mathcal{D}(p)} d, \qquad R_{\mathrm{span}}(p) = 2D_{\max}(p) + 1. \quad (9)$$

Quality and D14 both retain $D_{\max} = 4$, so D14 reduces the number of active branches from three to two while preserving the nine-row span that determines shared-buffer depth. Sections 4.3 and 4.5 describe the corresponding spatial lowering and programmable-logic realization. For compatibility and materialization analysis, FieldRT annotates every graph edge $e$ with tensor shape and precision, geometric semantics, frame dependence, consumer set, state lifetime, and materialization policy. These annotations determine whether a value crosses an external-memory boundary, remains in local storage, or flows directly between producers and consumers. For the set of cross-domain boundary edges $\mathcal{E}_{\mathrm{boundary}}$, the logical boundary volume is

$$T_{\mathrm{logical}} = \sum_{e \in \mathcal{E}_{\mathrm{boundary}}} \kappa_e B(e), \quad (10)$$

where $B(e)$ is the tensor size in bytes and $\kappa_e$ is the number of materialized crossings associated with edge $e$. This quantity captures the communication volume implied by the resolved graph partition and complements the board-level DMA measurements.

### 3.5 Backend Realization and Validation

The retained profiles are deployed as batch-one TensorRT engines on an AGX Orin through strict checkpoint loading, ONNX validation, and artifact-hashed engine construction. FP32 is the formal deployment condition; unrestricted FP16 is evaluated as a numerical stress condition. Successful engine construction is recorded separately from numerical validity. Each engine undergoes fixed-tensor finiteness and regression checks, followed by complete-test-set evaluation through the same decoder, rotated suppression, and evaluator as its PyTorch reference. We report accelerator-only, transfer-inclusive, and complete point-cloud-to-detection boundaries to determine how profile-level benefits survive the full pipeline. All comparisons use fixed batch size, power mode, clocks, CUDA Graph, synchronization, warm-up, and measurement settings. Exact interfaces, error metrics, execution boundaries, artifact lineage, and power protocols appear in Appendix I. FPGA realization and operation-matched validation are described in Sections 4.5.

## 4 FieldRT Hardware Co-Design

Our design jointly derives a geometry-preserving model profile and its hardware organization. For each candidate profile, FieldRT characterizes geometric function, numerical requirements, producer–consumer relationships, spatial reuse, and data lifetime. These properties form a geometric execution contract that guides complete-model TensorRT deployment on the NVIDIA Jetson AGX Orin and field-native streaming realization on the AMD Kria KV260.

The central design principle is that profile construction and hardware organization operate on the same geometric structure. A change in global-token capacity modifies both prediction context and reduction state. A change in axis-fusion behavior modifies both the algorithmic weighting rule and the fusion datapath. A change in the retained dilation set modifies both the multi-scale receptive field and the required line-buffer lifetime. FieldRT therefore treats model transformations as joint changes to prediction semantics, numerical state, communication, and storage.

### 4.1 Joint Profile and Backend Selection

Let $p \in \mathcal{P}_{\text{prof}}$ denote an AFN deployment profile and $b \in \mathcal{B}$ a realization backend. Our proposed FieldRT selects a profile–backend pair that minimizes realizable system cost while satisfying a vector of geometric and numerical quality requirements:

$$(p^\star, b^\star) = \mathop{\arg\min}_{p \in \mathcal{P}_{\text{prof}},\, b \in \mathcal{B}} C_{\text{sys}}(p, b) \quad \text{s.t.} \begin{cases} \mathbf{Q}_{\text{geom}}(p) \succeq \mathbf{Q}_{\min}, \\ V_{\text{num}}(p, b) = 1, \\ R(p, b) \preceq R_b^{\max}. \end{cases} \tag{11}$$

The quality vector

$$\mathbf{Q}(p, b) = \left[Q_{\text{dense}}, Q_{\text{inst}}, Q_{\text{strict}}, Q_{\text{yaw}}, Q_{\text{cross}}, Q_{\text{num}}\right]_{p,b} \tag{12}$$

captures dense recognition, instance detection, strict localization, orientation, cross-workload behavior, and deployment-time numerical validity. The system cost

$$C_{\text{sys}}(p, b) = \Phi_b\,(L, E, M, V, R) \tag{13}$$

aggregates the backend-relevant latency $L$, energy $E$, memory $M$, communication volume $V$, and hardware resources $R$. The backend-specific cost function $\Phi_b$ preserves the measurement boundary of each realization while providing a common profile-selection interface. The vector constraint preserves distinctions among perceptual objectives that respond variously to structural transformations. We evaluate changes to global context, prediction width, feature capacity, fusion, and multi-scale processing both independently and in combination. The analysis identifies profiles that preserve instance geometry under the complete configuration and target workload. A profile becomes eligible for realization when it satisfies three conditions: its geometric-quality vector meets the required operating point, its numerical format produces valid outputs on the target backend, and its resolved hardware configuration fits the available execution and storage resources. Profile eligibility is therefore determined jointly by prediction behavior and realizable execution.

### 4.2 Geometric Execution Contract

Figure 1 summarizes the FieldRT co-design workflow. Each profile is represented by a geometric execution contract

$$\Gamma(p) = [\mathcal{F}, \mathcal{E}, \mathcal{N}, \mathcal{T}, \mathcal{L}], \tag{14}$$

where $\mathcal{F}$ describes field semantics and spatial extent, $\mathcal{E}$ describes producer–consumer edges and reuse, $\mathcal{N}$ specifies numerical formats and accumulation requirements, $\mathcal{T}$ defines traversal and branch topology, and $\mathcal{L}$ describes state lifetime.

For each tensor edge $e = (u, v)$, the contract records

$$\gamma_e = [\mathbf{s}_e, \nu_e, \rho_e, \lambda_e, \eta_e], \tag{15}$$

where $\mathbf{s}_e$ is the tensor shape, $\nu_e$ is the numerical representation, $\rho_e$ describes the consumer and reuse pattern, $\lambda_e$ is the state lifetime, and $\eta_e$ records the geometric sensitivity associated with the edge. These descriptors distinguish tensors that have identical shapes but different geometric functions, consumers, or localization sensitivity.

The contract fabric resolves the profile and edge descriptors into a hardware configuration

$$\boldsymbol{\theta}_{\text{hw}} = \mathcal{R}_b\,(\Gamma(p)) = \left[\boldsymbol{\theta}_{\text{top}}, \boldsymbol{\theta}_{\text{flow}}, \boldsymbol{\theta}_{\text{num}}, \boldsymbol{\theta}_{\text{buf}}, \boldsymbol{\theta}_{\text{ctrl}}\right], \tag{16}$$

covering topology, dataflow, arithmetic format, buffering, and control. Profile variables determine global-token capacity, prediction width, feature width, active dilation branches, and fusion behavior. The resolved configuration determines enabled processing lanes, buffer dimensions, accumulator formats, stream routes, materialization boundaries, and address-generation parameters.

This mapping creates explicit correspondences between algorithm and hardware. The retained dilation set determines both receptive-field structure and buffer depth. Axis-fusion granularity determines both adaptive weighting capacity and gate-register organization. Feature and prediction widths determine both model capacity and channel parallelism. Global-token capacity determines both context representation and reduction-state storage.

Hardware execution returns:

$$\mathbf{r}_{\text{hw}} = [\text{cycles, bytes, occupancy, utilization, numeric status}] \tag{17}$$

through the contract interface. The returned measurements update system cost and profile eligibility. FieldRT therefore couples geometric sensitivity, configuration resolution, and measured deployment behavior in one profile selection loop.

### 4.3 Geometry-Native Spatial Lowering

Our design lowers recurring geometric structures into reusable spatial dataflow objects. The lowering preserves relationships among fields, consumers, neighborhoods, and aggregation state across the hardware mapping.

**Single-read field distribution.** AFN applies complementary axis-conditioned transformations to a common BEV field. A Multi-Field Engine reads each activation once and distributes the value to the $xy$, $xz$, and $yz$ consumers:

$$\mathrm{X}_t \longrightarrow \left\{\mathrm{X}_t^{xy}, \mathrm{X}_t^{xz}, \mathrm{X}_t^{yz}\right\}. \tag{18}$$

For $B$ consumers, independent materialization requires $B$ field reads, whereas the Multi-Field Engine reads the field once and performs on-chip distribution:

$$A_{\text{field}}^{\text{ind}} = B|\mathbf{X}|, \qquad A_{\text{field}}^{\text{shared}} = |\mathbf{X}|. \tag{19}$$

The fan-out maintains producer–consumer locality and supplies all consumers from shared field state. Axis-specific coordinate values and coefficients remain independent, preserving the geometric function of each branch.

**Coordinate-conditioned processing.** Each branch applies an axis-conditioned transformation using aligned activation, coordinate, and weight streams. A representative processing element

$$\mathbf{z}_{t,a} = \mathcal{A}\left(\mathbf{x}_t \odot \mathbf{c}_{t,a} \odot \mathbf{w}_a + \mathbf{b}_a\right), \tag{20}$$

where $a \in \{xy, xz, yz\}$, $\mathbf{c}_{t,a}$ is an axis-conditioned coordinate value, and $\mathcal{A}$ denotes the selected activation and accumulation convention. Separate activation, coordinate, and weight registers decouple stream arrival from wide accumulation while preserving alignment through the processing element.

**Streaming geometric fusion.** Axis-conditioned responses are combined directly in the stream:

$$\mathrm{S}_t = \sum_{a \in \{xy,xz,yz\}} \boldsymbol{\alpha}_{t,a} \odot \mathrm{X}_t^a, \tag{21}$$

where $\boldsymbol{\alpha}_{t,a}$ supports scalar or channel-wise and static or streamed weighting. The mode selects the granularity and source of the gate without changing the branch interface. Gate registers, multipliers, and an adder tree consume the axis-conditioned responses at their producer boundary, allowing the fused representation to continue without branch-wise external materialization.

**Lifetime-aware multi-dilation buffering.** Multi-dilation branches access overlapping neighborhoods of the same field. For active dilation set $\mathcal{D}(p)$, FieldRT retains one physical line buffer whose depth is determined by:

$$D_{\max}(p) = \max_{d \in \mathcal{D}(p)} d, \qquad R_{\text{span}}(p) = 2D_{\max}(p) + 1. \tag{22}$$

A window generator and selectable tap network expose the neighborhood required by each active branch:

$$\mathcal{W}_d(i,j) = \{X(i+rd, j+sd) \mid r,s \in \{-1,0,1\}\}. \tag{23}$$

All active branches share activation storage while retaining independent coefficient and accumulation paths. Profile changes update active taps and branch topology while preserving the common maximum-lifetime buffer. The memory organization therefore follows field lifetime instead of branch count.

**Streaming global aggregation.** Global context is represented as persistent reduction state. For field $\mathbf{X} \in \mathbb{R}^{C \times H \times W}$, the streaming reducer updates

$$g_c^{(t+1)} = g_c^{(t)} + x_{c,t}, \qquad c = 1, \ldots, C, \tag{24}$$

as each spatial sample enters the pipeline. The final state is normalized or projected into global tokens and forwarded to local–global prediction. Aggregation storage scales with channel and token capacity, rather full spatial field. Together, these objects map geometric structure into hardware organization: common fields determine fan-out, coordinate semantics determine conditioned processing, axis consumers determine fusion, dilation determines buffer lifetime, and global context determines reduction state.

### 4.4 Contract-Driven Specialization and Control

FieldRT separates compile-time capacity from execution-time mode selection. Compile-time parameters establish the maximum spatial and arithmetic envelope:

$$\boldsymbol{\theta}_{\text{compile}} = \left[C_{\max}, H, W, D_{\max}, P_{\text{ch}}, W_{\text{acc}}, M_{\text{gate}}\right], \tag{25}$$

where $P_{\text{ch}}$ is channel parallelism, $W_{\text{acc}}$ is accumulator width, and $M_{\text{gate}}$ denotes the supported fusion-mode set. Runtime configuration selects an admissible subconfiguration:

$$\boldsymbol{\theta}_{\text{run}} = \left[\mathbf{m}_{\text{branch}}, \mathbf{m}_{\text{lane}}, m_{\text{gate}}, m_{\text{precision}}, \mathbf{a}_{\text{base}}\right], \tag{26}$$

where the masks activate retained branches and lanes, and $\mathbf{a}_{\text{base}}$ supplies the resolved memory addresses.

The control plane contains profile registers, a configuration decoder, loop-and-address controllers, and status counters. Profile registers encode bounds, supported numerical formats, and the active profile mode. The decoder converts these fields into branch enables, lane masks, bypass signals, tap selection, and fusion control. Loop controllers generate spatial and channel iteration, DMA descriptor sequences, and address progression. Status counters return completed cycles, transferred bytes, occupancy, resource state, and numerical flags. This division preserves a specialized datapath while allowing multiple qualified profiles to share one parameterized architecture. Compile-time capacity guarantees bounded storage and arithmetic structure, while runtime selection activates the profile admitted by the resolved execution contract.

### 4.5 Programmable-Logic Realization

The KV260 realization comprises parameterized HLS kernels connected through AXI4-Stream interfaces and configured through AXI-Lite registers. MM2S DMA channels supply activations, coordinates, and coefficients, while dedicated S2MM channels drain materialized outputs. Local SRAM and register storage retain reusable coordinate, coefficient, and activation state. Elastic FIFOs decouple adjacent processing stages and absorb variation in stream readiness.

For multi-dilation execution, weights and activations enter through explicit transfer phases before the shared-buffer kernel processes the activation stream:

$$\begin{aligned} &\text{MM2S} \rightarrow \text{controlled demultiplexing} \rightarrow \text{shared-buffer kernel} \\ &\quad \rightarrow \text{elastic streams} \rightarrow \text{S2MM} \rightarrow \text{DDR}. \end{aligned} \tag{27}$$

The demultiplexer associates each transfer phase with a destination and a logical beat count. For descriptor sequence $\{n_1, \ldots, n_J\}$, the logical tensor boundary is

$$N_{\text{tensor}} = \sum_{j=1}^{J} n_j. \tag{28}$$

The kernel advances according to $N_{\text{tensor}}$, preserving one tensor transaction across several physical DMA descriptors. Dedicated output channels drain each materialized branch continuously, while elastic stream stages isolate the compute pipeline from output acceptance. The MFE and fusion datapaths form a direct producer–consumer stream. The MFE generates aligned axis-conditioned streams from one field traversal, and fusion consumes these streams under the selected gate mode. The global reduction datapath maintains channel state during the traversal and emits one bounded output vector. These connections retain intermediate values within the spatial pipeline and expose external memory boundaries only where the resolved contract requires materialization.

## 4.6 Cost and Numerical Feedback

A resolved profile produces both an execution configuration and an expected measurement schema. FieldRT collects hardware observations as

$$\mathbf{c}_{\text{hw}} = [\ell_{\text{txn}}, B_{\text{in}}, B_{\text{out}}, U_{\text{LUT}}, U_{\text{FF}}, U_{\text{DSP}}, U_{\text{BRAM}}], \tag{29}$$

where $\ell_{\text{txn}}$ is transaction latency, $B_{\text{in}}$ and $B_{\text{out}}$ describe materialized traffic, and $U_{\cdot}$ denotes resource utilization. These quantities update $C_{\text{sys}}(p, b)$ and expose whether profile structure is limited by arithmetic, storage, or communication.

FieldRT qualifies each hardware operation against an operation-matched software reference. For output $\mathbf{y}$ and reference $\mathbf{y}^{\text{ref}}$, qualification reports:

$$N_{\neq} = \sum_{k=1}^{K} \mathbf{1}\left[y_k \neq y_k^{\text{ref}}\right], \qquad \text{MAE} = \frac{1}{K}\sum_{k=1}^{K}\left|y_k - y_k^{\text{ref}}\right|. \tag{30}$$

Integer execution achieves bit-exact agreement when $N_{\neq} = 0$ and MAE $= 0$ over the complete output tensor. The qualification chain compares the operation-matched software reference, HLS C simulation, and physical-board output. Tensor identities, numerical formats, and bitstream identities bind the three stages to the same execution contract. Complete TensorRT profiles use the corresponding model-level feedback path. Full-workload execution contributes finite-output status, task-level agreement, latency, and energy to $\mathbf{Q}(p, b)$ and $C_{\text{sys}}(p, b)$. The Contract Fabric combines these backend-specific measurements through profile cost and eligibility, completing the FieldRT co-design loop. Appendix L reports the complete HLS kernel matrix, resource and initiation-interval sweeps, board transactions, DMA accounting, compatibility-path partitioning, operation-matched validation, bitstream reuse, power characterization, and platform-integration analysis.

# 5 General Evaluation

The evaluation is organized in the same order as the development of FieldRT. An initial cross-model comparison establishes AFN as the task-model foundation before any hardware-oriented transformation is introduced. The subsequent experiments therefore remain within the AFN family and progressively evaluate isolated structural sensitivity, conditional interactions among transformations, cross-workload profile stability, and the geometric quality of the resulting deployment profiles. After selecting the balanced profile, we evaluate whether the measured profile behavior survives backend conversion and whole-model deployment. PyTorch-to-TensorRT regression verifies numerical fidelity, and paired accelerator-only and point-cloud-to-detection measurements characterize latency, throughput, and energy under explicitly separated system boundaries. The execution-contract realizations and FPGA compatibility-path experiments are reported separately in Section 6.

## 5.1 Experimental Setup

**Workloads.** We evaluate oriented-BEV perception on Simulation and Real-world point clouds collection (shown in Figure. 12). Their spatial ranges, class sets, coordinate conventions, and intensity availability remain workload-specific. Complete preprocessing parameters, class lists, evaluation thresholds, and split identities appear in Appendix A.

**Evaluation protocol.** We report oriented-box $\text{mAP}_{0.50:0.95}$, $\text{AP}_{50}$, $\text{AP}_{75}$, mIoU, Dense-mAP, and median $\pi$-periodic yaw error. Dense-mAP evaluates the categorical field before final instance formation, whereas the instance metrics additionally depend on peak extraction, object separation, box regression, candidate ordering, and rotated post-processing. Besides, figure 13 compares oriented ground-truth and predicted boxes for representative point-cloud frames. The examples cover different scene densities, object counts, spatial distributions, and sensing ranges.

## 5.2 Co-Design Foundation

FieldRT follows a two-stage evaluation strategy. We first select the task-model foundation by comparing representative perception model families under the established baseline protocol. Only after AFN is selected do we introduce the stricter oriented-instance protocol used to analyze AFN structures, construct deployment profiles, and evaluate their TensorRT, AGX, and FPGA realizations.

Fig.3 reports the initial model-selection results on Simulation, Real, and LiDAR-Net[11]. The comparison includes PillarNet, PointNet++, BEV-MAE, GroupFree3D, and AFN. This stage evaluates aggregate classification, detection, and localization behavior using the metrics available in the original model-selection benchmark.

The purpose of this first stage is model-family selection rather than deployment-profile selection. After selecting AFN, all subsequent FieldRT experiments remain within the AFN family and use the stricter matched oriented-instance evaluator. This separation allows the co-design study to attribute quality, numerical, and deployment differences to controlled AFN transformations rather than to unrelated differences among model families. The remaining experiments consequently operate within AFN. They use the formal FieldRT evaluator to distinguish permissive from strict overlap, dense categorical quality from final instance quality, and isolated structural effects from complete-profile behavior. The AGX and FPGA studies are therefore downstream realizations of a task model selected before hardware-specific optimization begins.

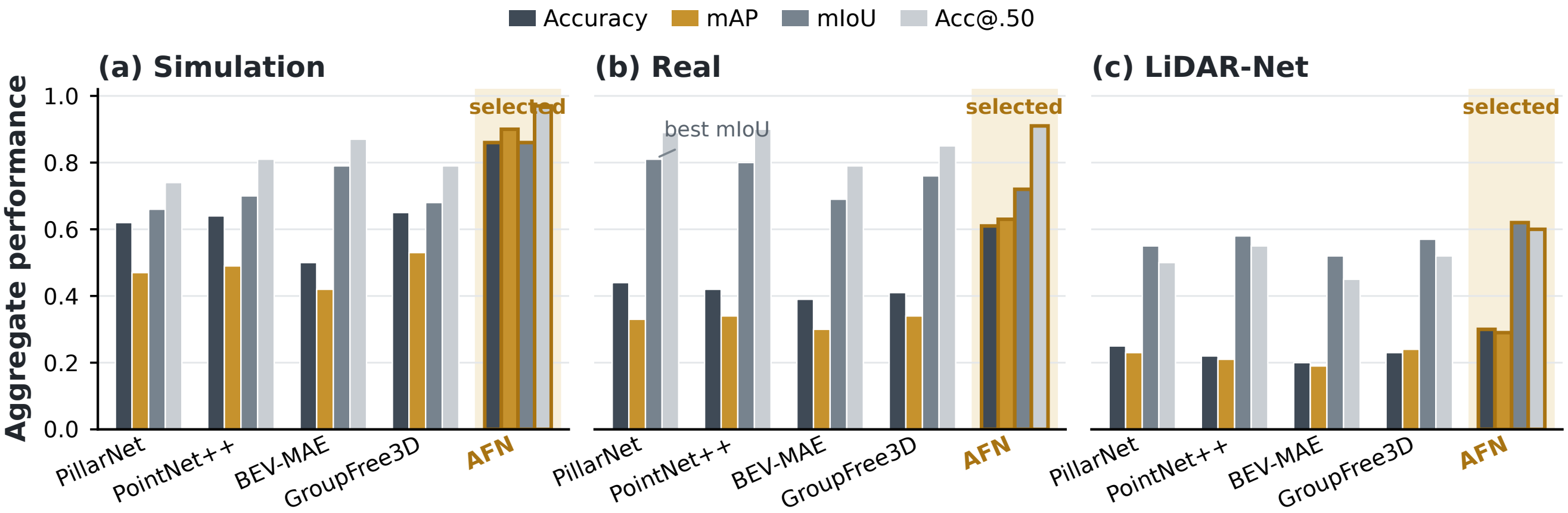

**Figure 3: Initial task-model comparison under the established baseline protocol. AFN provides the strongest aggregate performance across the three workloads, except for Real mIoU.**

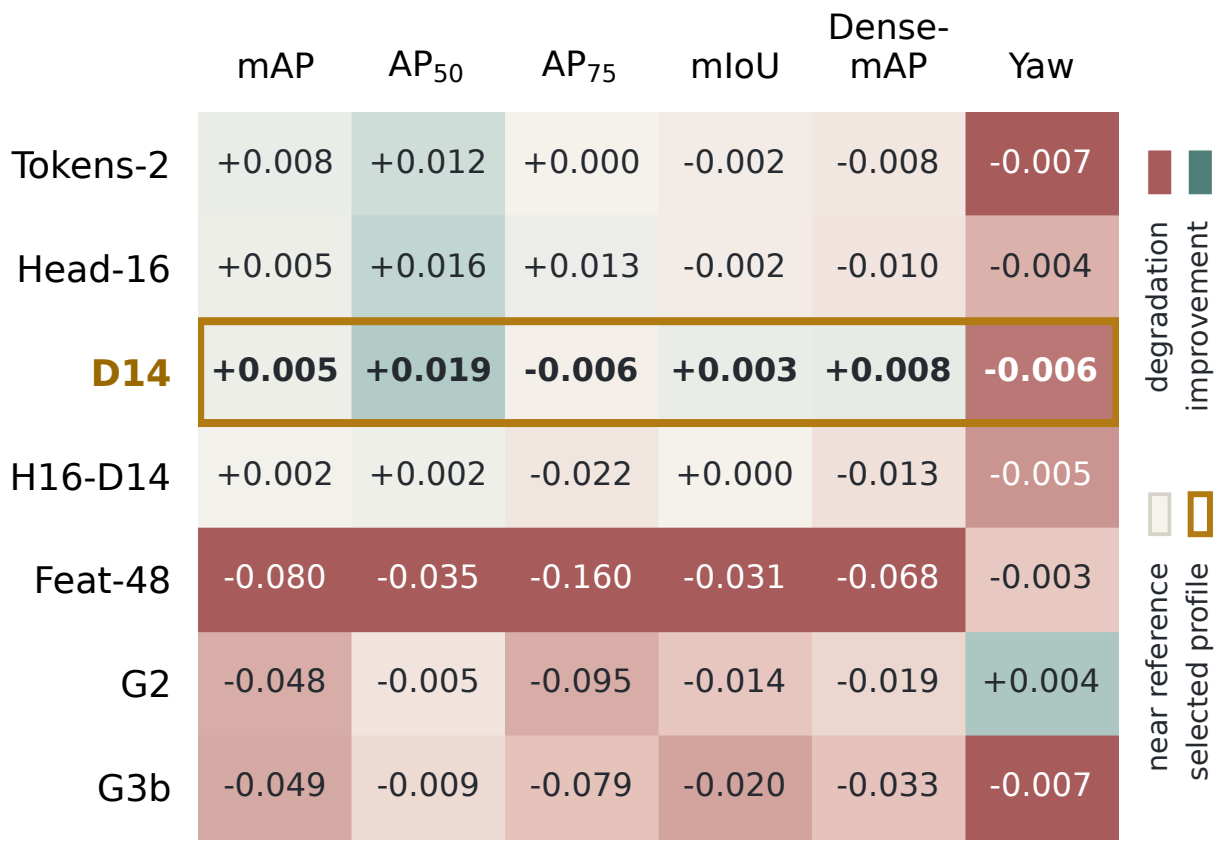

**Figure 4: AFN profile sensitivity relative to Quality. Cell values report absolute metric changes, and D14 is outlined.**

## 5.3 Geometric Sensitivity of AFN Structures

We evaluate isolated transformations of the global-token count, prediction-head width, dilation set, common feature width, and geometric-axis fusion mechanism. H16-D14 is additionally included to expose transformation composition. Figure 4 reports the resulting metric changes relative to Quality. Complete results in Appendix F.

**Permissive detection does not determine strict geometry.** G2 replaces input-dependent fusion with static channel-wise coefficients. The resulting gap between permissive and strict localization indicates that sample adaptivity is important for preserving high-overlap geometry. Restoring input dependence in G3a partially recovers this loss, whereas the finer sample-channel formulation in G3b provides no further overall benefit. Input dependence therefore matters, but increasing coefficient granularity alone is insufficient to capture branch-specific geometric reliability.

**Uniform width reduction is not equivalent to selective structural removal.** Uniform feature-width reduction is substantially more damaging than the selective transformations in Tokens-2, Head-16, and D14, indicating that AFN redundancy is structurally non-uniform and cannot be identified by uniform pruning.

## 5.4 Conditional Geometric Redundancy

The effect of a structural transformation depends on the metric, workload, and surrounding profile. Figure 5 shows that profile rankings change between dense and instance quality, between $AP_{50}$ and $AP_{75}$, and between Simulation and Real. E.g. Tokens-2 improves Real Dense-mAP from 0.460 to 0.490 but reduces instance mAP from 0.106 to 0.069, demonstrating that dense recognition does not determine instance quality. D14 remains comparatively stable across workloads, whereas Tokens-2 and H16-D14 lose substantial Real instance quality. The negative interaction residuals of H16-D14 further show individually plausible transformations need not compose safely. Deployment-relevant redundancy is therefore a property of the complete profile under a particular workload and evaluation, rather than an intrinsic property of an isolated component.

## 5.5 Cross-Workload Quality of the Profiles

We next evaluate the five formal deployment profiles through complete-test TensorRT FP32 inference. Figure 6 summarizes their geometric quality relative to the workload-matched Quality profile. The profiles remain closely grouped on Simulation, with different transformations leading individual metrics. Their Real behavior is more discriminative. Tokens-2, Head-16, and H16-D14 exhibit substantial losses in instance quality, whereas D14 remains closest to Quality across the reported metrics and improves strict localization. D14 removes only the intermediate $d = 2$ denoising branch while retaining the Quality feature width, adaptive fusion, global-token capacity, prediction-head width, and maximum dilation. This selective transformation reduces computation without uniformly narrowing the geometric representation or jointly weakening multiple compensating structures.

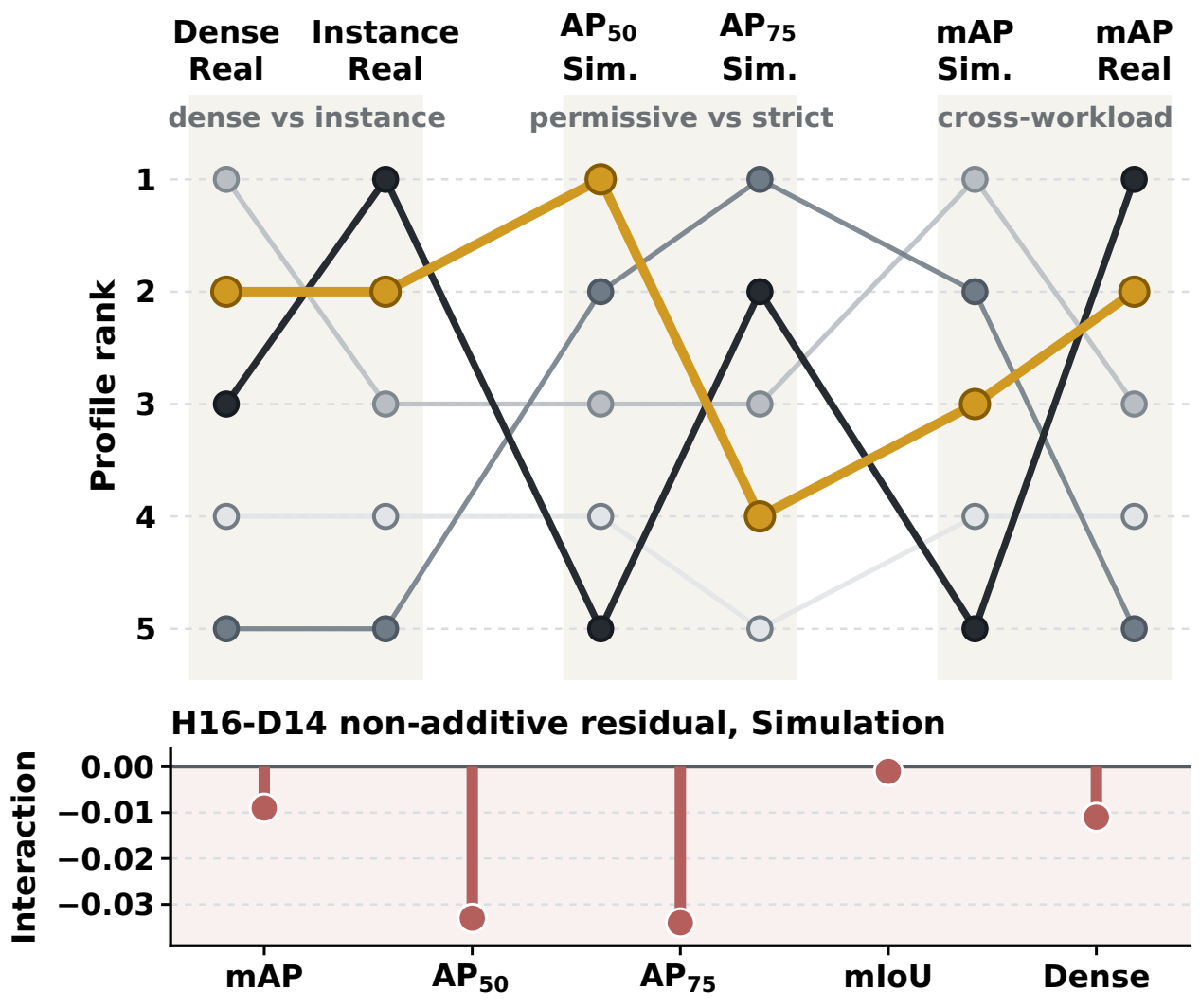

Figure 5: Conditional geometric redundancy. Profile rankings vary across metrics and workloads.

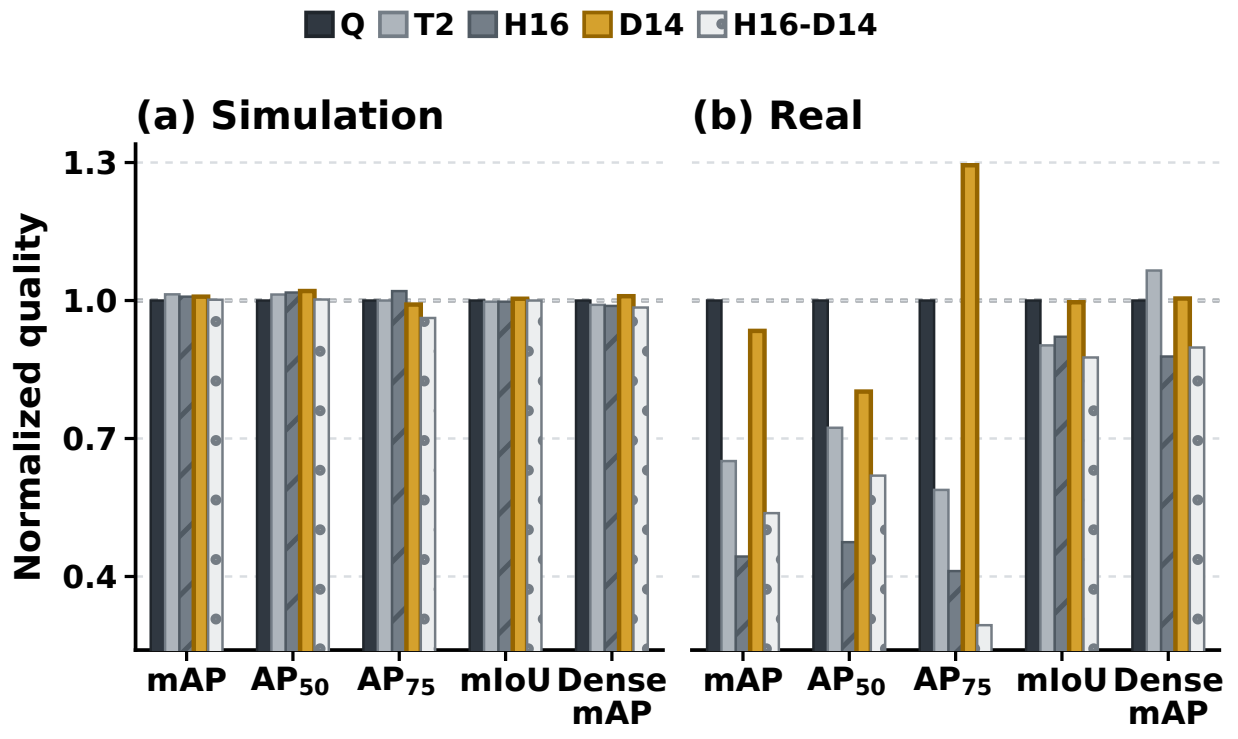

Figure 6: TensorRT quality to workload-matched Quality profile. D14 provides the strongest cross-workload balance.

## 5.6 TensorRT Numerical Qualification

All FP32 profile–workload engines complete their corresponding test sets with finite classification and regression outputs. Across the eight non-reference candidate comparisons, five satisfy the registered strict criterion and the remaining three satisfy complete-test metric regression; none fails. The largest observed core-metric deviation is 0.003104, showing that the profile-quality conclusions are preserved through ONNX conversion and TensorRT optimization. Complete per-profile and per-metric regression results appear in Appendix H. Compilation alone, however, does not establish numerical validity. The unrestricted FP16 engine builds successfully but produces non-finite classification and regression outputs on genuine Simulation inputs, while the corresponding PyTorch and TensorRT FP32 paths remain finite. A partially constrained mixed engine passes layer audit and short execution, but still contains substantial internal FP16 computation and has not completed genuine-input task-level regression. We therefore qualify only the FP32 engines and do not treat successful compilation or FP32-formatted outputs as evidence of valid low-precision deployment.

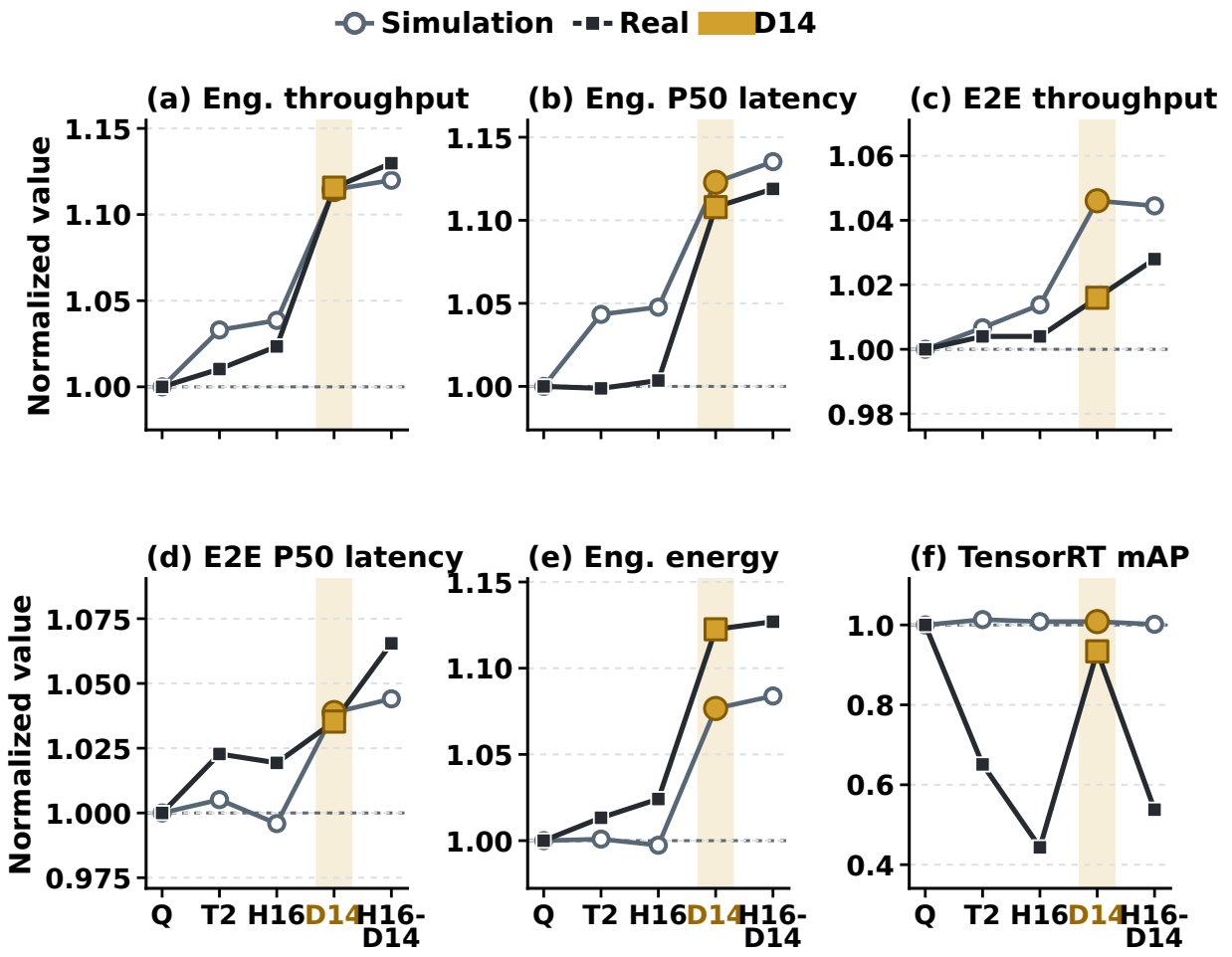

Figure 7: AGX deployment results normalized to workload-matched Quality. Higher values are preferable in every panel.

## 5.7 AGX Deployment

We evaluate all profiles under accelerator-only and complete point-cloud-to-detection boundaries. Figure 7 summarizes accelerator performance, end-to-end efficiency, engine-only energy, and TensorRT task quality relative to the workload-matched Quality profile. Energy is measured separately under paired engine-only execution.

**Accelerator-only behavior.** D14 and H16-D14 provide the clearest accelerator-level improvements on both workloads, whereas Tokens-2 and Head-16 yield comparatively limited benefit, particularly on Real. The small latency gap between D14 and H16-D14 indicates that most of the measured gain is already obtained by removing the intermediate denoising branch, with little additional benefit from the accompanying head reduction. **Complete-deployment behavior.** The accelerator-level gains are attenuated under the complete point-cloud-to-detection boundary because preprocessing, communication, output handling, and geometric post-processing remain largely unaffected by the profile transformation. H16-D14 achieves the lowest end-to-end latency, particularly on Real, but its small additional speed advantage over D14 is accompanied by a substantial loss in instance quality. Minimum latency therefore does not identify the strongest complete deployment profile. **Engine-only energy.** D14 and H16-D14 occupy the most efficient measured engine-energy tier on both workloads. H16-D14 provides only a marginal additional energy benefit over D14 while substantially degrading Real instance quality. D14 therefore captures most of the available engine-energy reduction without incurring the quality loss associated with the more aggressive combined profile.

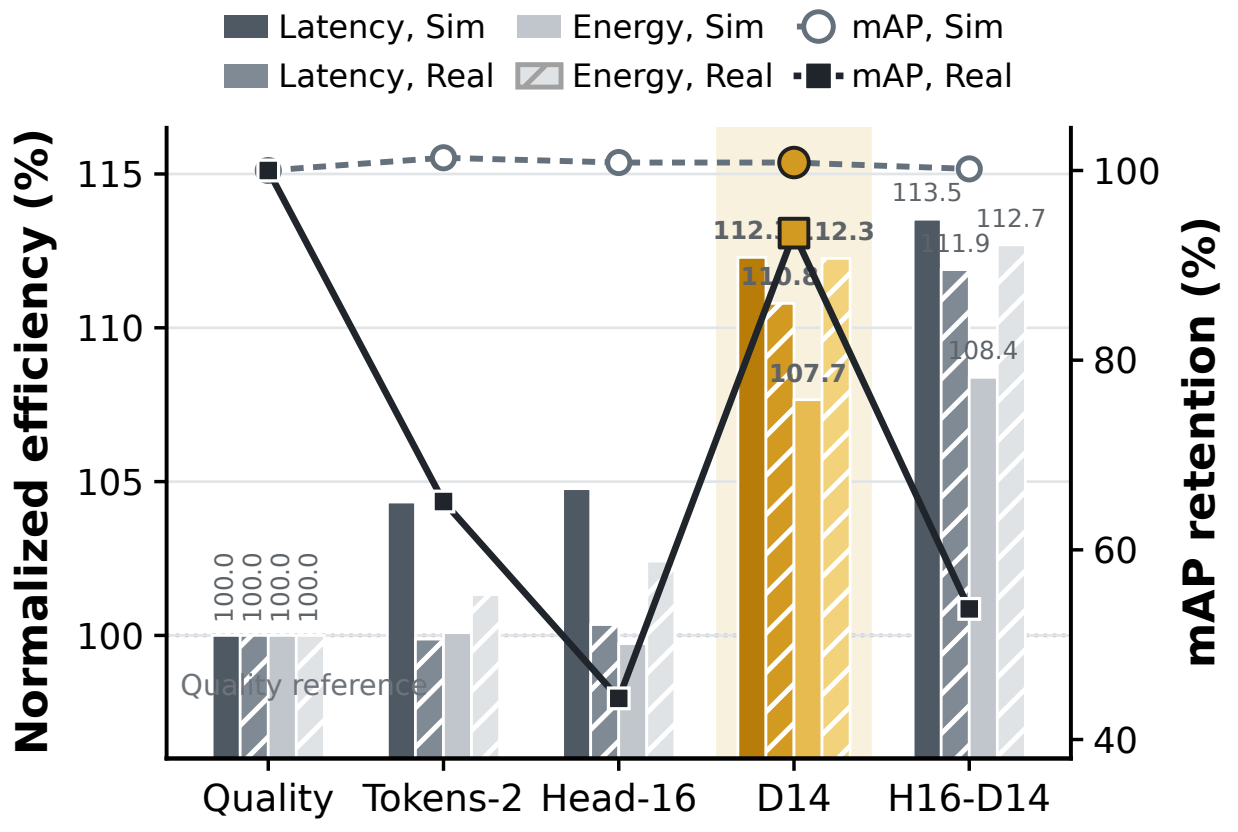

**Figure 8: Normalized profile quality and efficiency. D14 captures most of the efficiency gain while maintaining the strongest cross-workload quality balance.**

## 5.8 Deployment-Profile Selection

Figure 8 summarizes profile quality and efficiency relative to the workload-matched Quality reference. Tokens-2 and Head-16 do not preserve Real instance quality, while H16-D14 provides only a marginal efficiency advantage over D14 at a substantial geometric-quality cost. D14 instead removes only the intermediate $d = 2$ denoising branch, retaining the remaining Quality structures and most of the available latency and engine-energy gains. We therefore select D14 as the deployment profile based on its cross-workload quality, TensorRT numerical validity, and deployment efficiency.

# 6 Field-Aware FPGA Realization and Evaluation

We evaluate FieldRT's profile-guided execution contracts on the AMD Kria KV260. The evaluation characterizes compatibility-path fragmentation, shared-buffer resource feasibility, production-scale numerical preservation, and board-level data movement. Together, these experiments connect AFN profile selection with the realizability of geometry-native programmable-logic execution. Figure 2 shows the implemented KV260 used for custom-PL integration.

## 6.1 Compatibility-Path Fragmentation

The standard Vitis-AI compatibility flow partitions both compiled AFN profiles into alternating DPU and CPU-fallback regions, as summarized in Figure 9. The substantial fallback fraction shows that graph compilation alone does not provide an effective geometry-native execution path. Unsupported geometric and statistical operations repeatedly cross the accelerator boundary, motivating the custom-PL contracts evaluated below.

## 6.2 Shared-Buffer Multi-Dilation Execution

A literal hardware mapping allocates an independent line buffer to each dilation branch. This requires 408 $BRAM_{18K}$ blocks for the three-branch Tokens-2 profile and 288 blocks for the two-branch D14 family, corresponding to 142% and 100% of the KV260 budget, respectively. FieldRT instead retains the common input field in one line buffer sized by the maximum active dilation. Dilation-specific branches reuse this buffer through distinct tap schedules. The resulting production-scale Multi2 and Multi3 implementations each use 216 $BRAM_{18K}$ blocks, reducing the memory footprint to 75% of device capacity and making the three-branch realizable.

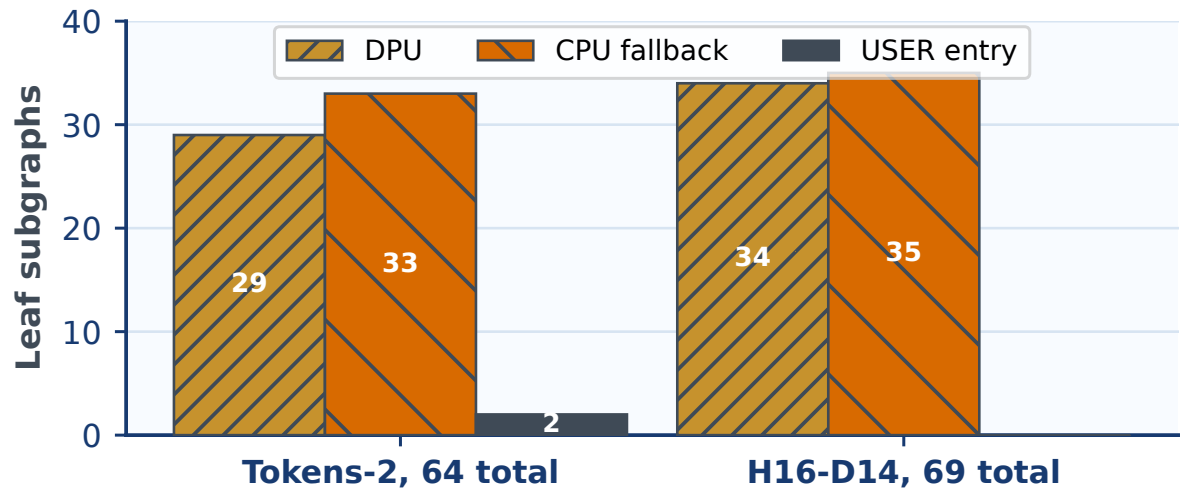

**Figure 9: Partition under the Vitis-AI compatibility flow. Both AFN profiles retain substantial CPU-fallback execution, motivating geometry-native programmable-logic contracts.**

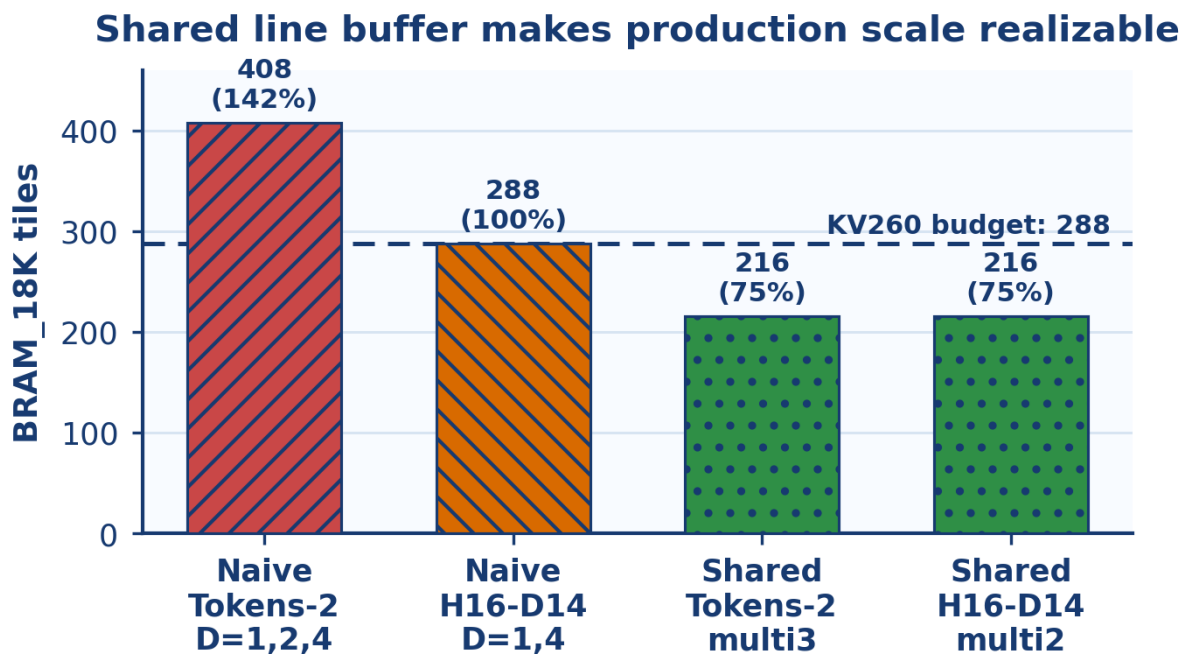

**Figure 10: $BRAM_{18K}$ requirements of per-dilation buffers and the shared-buffer implementation. Shared retention reduces Tokens-2 from 408 to 216 blocks and the D14 family from 288 to 216 blocks against the 288-block KV260 budget.**

## 6.3 Synthesis Characteristics

Figure 11 summarizes the production-scale synthesis characteristics of the implemented custom-PL kernels. MFE, fusion, and reduction provide lightweight streaming primitives, while multi-dilation denoising accounts for the dominant on-chip memory demand. Multi2 and Multi3 use 75% of the available BRAM but remain comparatively modest in logic and DSP utilization.

## 6.4 Production-Scale Board Validation

We validate the shared-buffer two-branch kernel using complete $64 \times 400 \times 400$ checkpoint-derived activations and weights. The signed 8-bit input contains 10,240,000 elements, and each D1 and D4 branch produces 10,240,000 signed 32-bit outputs, corresponding to 40.96 MB per branch. For both Simulation and Real fixtures, KV260 outputs are bit exact against their operation-matched CPU references. Across the four fixture–branch comparisons, all 10,240,000 elements per branch match exactly, with zero mean absolute error, zero maximum absolute error, and cosine similarity 1.0. The same frozen bitstream is reused across the two checkpoint-derived fixtures, demonstrating production-scale numerical preservation across both workloads. The production-scale MFE–fusion path provides complementary validation of FieldRT's streaming execution contracts. It processes a $64 \times 400 \times 400$ field with zero mismatches over 10,240,000 output elements and produces byte-identical results across four consecutive board invocations. This experiment exercises the static-scalar gate at production scale. Static-channel, sample-scalar, and sample-channel gating are validated separately at tile scale. Streaming global reduction provides a third operation-matched validation path. Its CPU reference, HLS C simulation, and KV260 board output agree bit exactly over all 32 reduced channel values, with zero mean absolute error and cosine similarity 1.0.

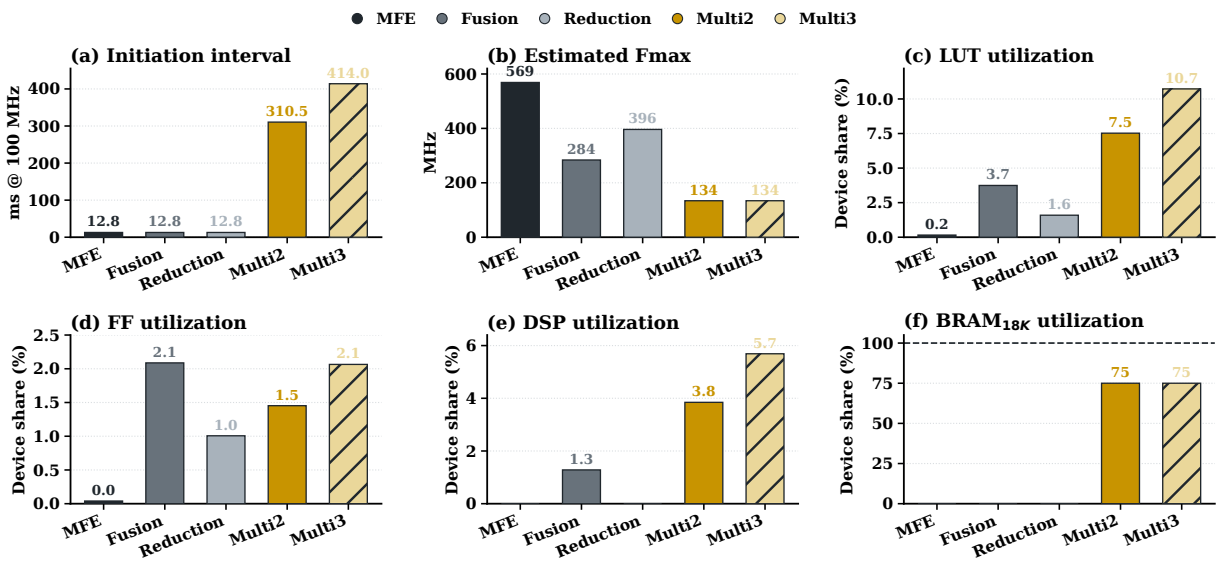

**Figure 11: Production-scale HLS characteristics of the custom-PL kernels. The panels report initiation interval at 100 MHz, HLS-estimated Fmax, and LUT, FF, DSP, and $BRAM_{18K}$ utilization. Zero-valued entries are retained to preserve a consistent kernel ordering across panels. Fmax values are derived from HLS synthesis estimates.**

## 6.5 Transaction and Data-Delivery Bottleneck

The Real production-scale transaction uses one MM2S input path and two dedicated S2MM output paths. The 10.24-MB activation is divided into 626 descriptors by the current 16,383-byte descriptor limit. All three DMA channels complete successfully, and each output path reports the expected 40.96-MB branch output.

The independently measured DMA-inclusive accelerator boundary is 441.953 ms. Weight loading requires 0.143 ms before this boundary, while host-side output dumping is excluded. Within the transaction, the input-DMA diagnostic interval is 429.047 ms and kernel-plus-S2MM completion requires 3.052 ms. These diagnostic intervals are nested measurements and are not an exhaustive additive decomposition of the outer interval.

The transaction is therefore data-delivery dominated rather than arithmetic dominated. The descriptor-limited input transfer accounts for most of the observed board time, identifying descriptor aggregation, larger contiguous transfers, and direct producer-to-kernel streaming as the next optimization targets.

## 7 Conclusion

We presented FieldRT, a geometry-driven algorithm-hardware co-design for edge device perception. FieldRT jointly derives a geometry-preserving deployment profile and its spatial execution architecture from a common analysis of prediction sensitivity, numerical representation, producer–consumer structure, communication, and data lifetime. Controlled transformations of AFN revealed that deployment-relevant redundancy is conditional rather than uniform. Instance detection, strict localization, orientation, and cross-workload robustness respond variously to structural changes, while transformations that are effective independently can interact negatively when combined. A reliable hardware profile must therefore be selected under the complete model configuration and target workload rather than through uniform compression, aggregate accuracy, or isolated complexity proxies. FieldRT translates the retained geometric structure into single-read field distribution, streaming fusion, lifetime-aware multi-dilation buffering, and streaming reduction. Shared spatial state reduces the three-branch buffering requirement from 408 to 216 $BRAM_{18K}$ blocks, converting an infeasible organization into a realizable design. Production-scale execution on programmable logic establishes numerical preservation of the resulting spatial contracts, while complete TensorRT deployment demonstrates how geometry-sensitive profile choices affect task quality, numerical validity, latency, power, and energy at the system boundary. Together, the results show that model compactness, operator compatibility, and backend acceptance do not independently establish effective deployment. The central design object is the relationship between the algorithmic structures that preserve instance geometry and the numerical state, storage lifetime, communication, and dataflow through which those structures are executed. FieldRT makes this relationship explicit, providing a unified basis for deciding both what an edge perception system must preserve and how the retained computation should be realized.

**Table 2: Production-scale Real board transaction.**

| Measurement | Value |
|---|---|
| Input activation | 10.24 MB |
| Input descriptors | 626 |
| Descriptor payload limit | 16,383 bytes |
| Output per branch | 40.96 MB |
| Weight load, before boundary | 0.143 ms |
| Input-DMA diagnostic interval | 429.047 ms |
| Kernel and S2MM diagnostic interval | 3.052 ms |
| DMA-inclusive accelerator boundary | 441.953 ms |
| Host output dump, outside boundary | 1,574.024 ms |

## Acknowledgments

This work was funded by the European Commission's Marie Skłodowska-Curie Action (MSCA) Project RAICAM (GA101072634).

**Table 3: Complete Simulation evaluation protocol.**

| Item | Configuration |
|---|---|
| Train/validation/test samples | 2,040 / 255 / 255 |
| Foreground classes | Table, Chair, Shelf, Box, Stair |
| BEV dimensions | 400 × 400 |
| $x$ range | $[-10, 10]$ m |
| $y$ range | $[-10, 10]$ m |
| $z$ range | $[-1, 2]$ m |
| Input channels | 6 |
| Intensity | Disabled |
| Coordinate convention | LiDAR |
| Orientation encoding | Double angle |
| Regression channels | 6 |
| Score threshold | 0.30 |
| NMS IoU threshold | 0.20 |

**Table 4: Complete Real evaluation protocol.**

| Item | Configuration |
|---|---|
| Train/validation/test samples | 416 / 52 / 52 |
| Foreground classes | Couch, Table, Person, Chair |
| BEV dimensions | 400 × 400 |
| $x$ range | $[-2, 2]$ m |
| $y$ range | $[-2, 2]$ m |
| $z$ range | $[-1, 2]$ m |
| Input channels | 6 |
| Intensity | Enabled |
| Coordinate convention | KITTI camera |
| Orientation encoding | Double angle |
| Regression channels | 6 |
| Score threshold | 0.30 |
| NMS IoU threshold | 0.20 |

# A Appendix

# A Complete Evaluation Protocol

This appendix specifies the workload interfaces, evaluation metrics, model configuration, TensorRT qualification procedure, deployment boundaries, power accounting, and artifact-lineage rules used by FieldRT. The appendix separates PyTorch task quality, TensorRT numerical validity, accelerator-only execution, transfer-inclusive execution, evaluator-path timing, inference-only deployment, and hardware-compatibility analysis. Results from these boundaries are not interchanged.

## A.1 Simulation Workload

Simulation point clouds are loaded from the registered `.bin` or `.npy` sources. Non-finite points and points outside the registered three-dimensional region of interest are discarded before BEV aggregation. Fixed-length point sampling is disabled, so every finite in-range point contributes to the BEV statistics.

## A.2 Real Workload

The Real `.bin` files are direct variable-length Livox point-cloud frames. The formal loader does not pad, duplicate, or resample these frames to a fixed number of points. It removes non-finite and out-of-range points and then aggregates all remaining points into the BEV grid. For a KITTI-camera location $(x_{\mathrm{cam}}, y_{\mathrm{cam}}, z_{\mathrm{cam}})$, the Real data loader constructs the AFN coordinates as

$$x_{\mathrm{AFN}} = z_{\mathrm{cam}}, \qquad y_{\mathrm{AFN}} = -x_{\mathrm{cam}}, \qquad z_{\mathrm{AFN}} = y_{\mathrm{cam}}. \tag{31}$$

The corresponding AFN heading is

$$\theta_{\mathrm{AFN}} = -r_y - \frac{\pi}{2}, \tag{32}$$

followed by wrapping to $[-\pi, \pi)$.

## A.3 Evaluation Metrics

The complete metric family contains:

- oriented-box mAP averaged over IoU thresholds $0.50 : 0.05 : 0.95$;
- $\mathrm{AP}_{50}$ and $\mathrm{AP}_{75}$;
- class-correct best-match mIoU, with unmatched ground-truth instances assigned zero;
- Acc@IoU0.50 and Acc@IoU0.75;
- per-class Dense AP and Dense-mAP;
- foreground classification accuracy and macro-F1;
- center, length, width, and $\pi$-periodic yaw errors; and
- oracle localization using ground-truth center, size, or yaw.

Instance AP uses score-sorted, class-specific one-to-one matching within each frame. All principal manuscript tables display values rounded to three decimal places. Percentage changes, normalized changes, and backend differences are calculated from the unrounded source results.

# B Complete Task-Model Benchmark

Before constructing deployment profiles, AFN is compared with PillarNet, PointNet++, BEV-MAE, and GroupFree3D under the established LUMI task-model evaluation protocol. The comparison covers the Simulation, Real, and LiDAR-Net workloads and provides the task-model evidence used to motivate the subsequent AFN profile analysis. These results are reported separately from the later controlled profile, TensorRT, and AGX Orin evaluations. The archived source table contains one additional unlabeled score column; that column is omitted here until its metric definition is recovered.

# C Complete AFN Architecture and Training Protocol

This section provides the architecture and training details omitted from the main text. The main paper retains only the operations required to explain geometric sensitivity and the execution contract.

## C.1 BEV Construction

For frame $t$, each point is represented as

$$\mathbf{p}_{t,n} = (x_{t,n}, y_{t,n}, z_{t,n}, r_{t,n}), \tag{33}$$

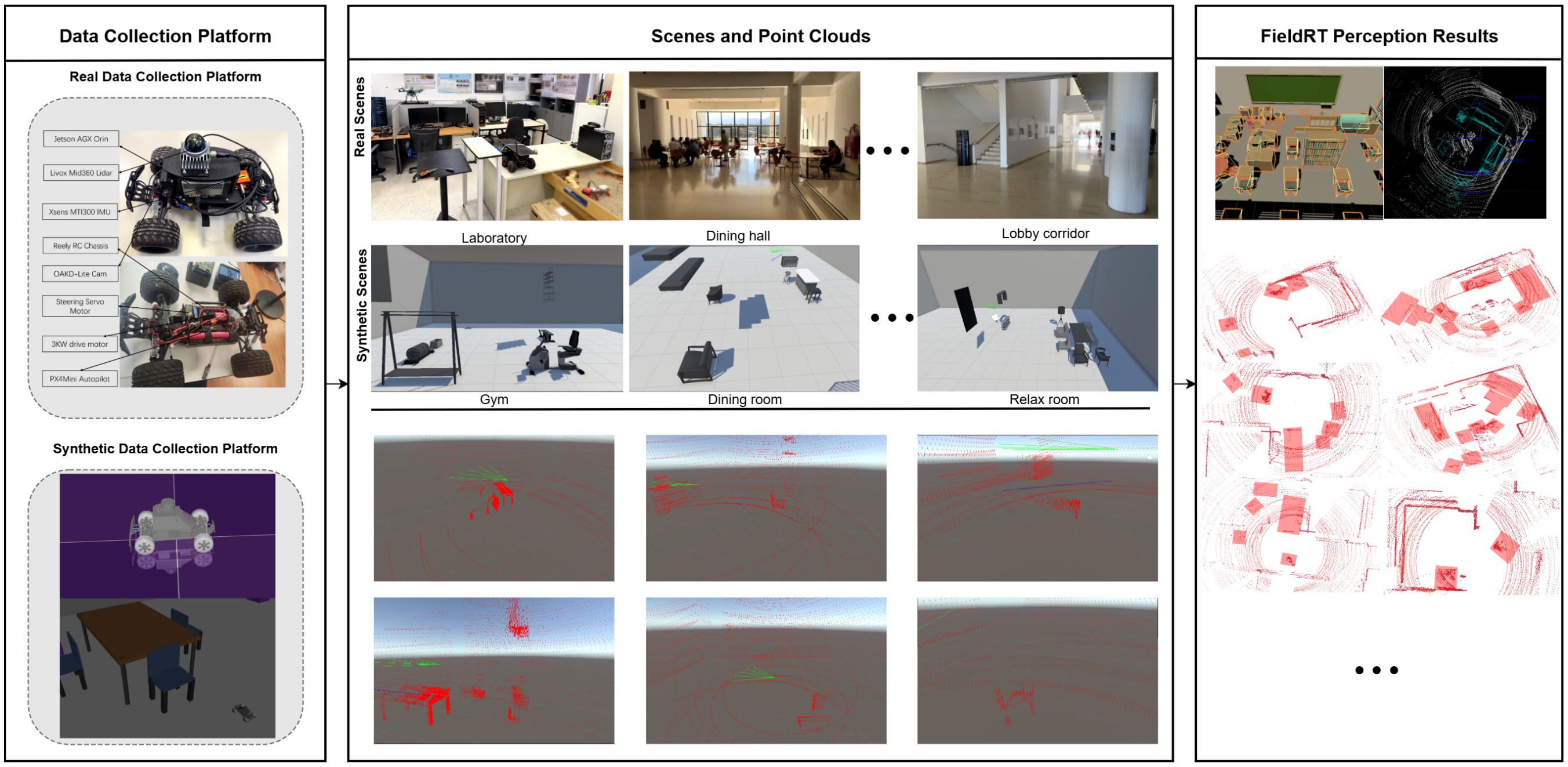

Figure 12: FieldRT perception workloads and data-collection platforms. Real point clouds are collected with the instrumented mobile platform across indoor laboratory, dining, and corridor environments, while synthetic scenes provide complementary layouts and object configurations. The collected point clouds form the Simulation and Real workloads used to construct and evaluate AFN deployment profiles, with representative perception outputs shown on the right.

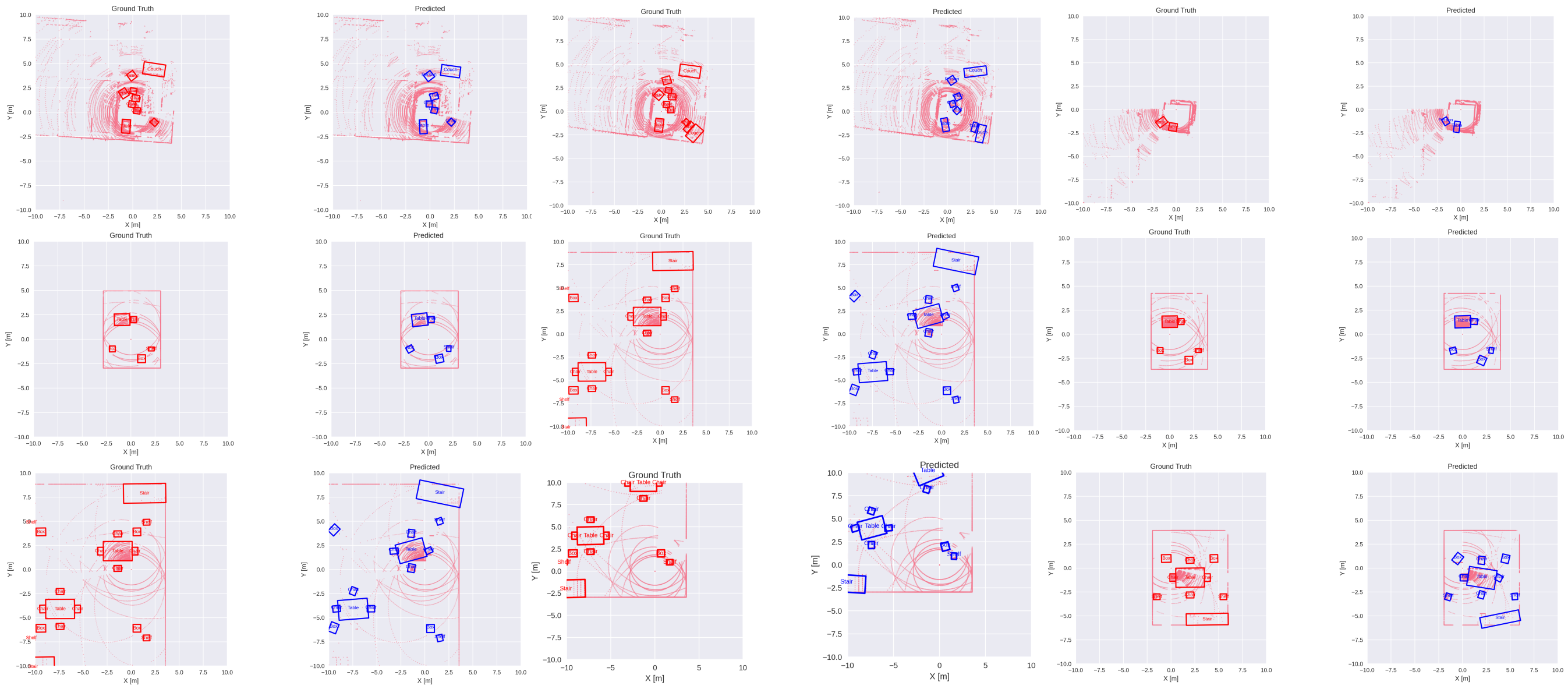

Figure 13: Representative oriented-detection results. Each adjacent pair shows the ground-truth annotations and AFN predictions for the same point-cloud frame. Red boxes denote ground-truth objects and blue boxes denote predicted objects. The examples cover sparse and crowded scenes, near- and long-range objects, and varied object positions and orientations.

where $r_{t,n}$ denotes intensity when available. Points outside the workload-specific region of interest are removed. The retained coordinates are mapped to BEV indices

$$u_{t,n} = \left\lfloor \frac{x_{t,n} - x_{\min}}{x_{\max} - x_{\min}} W \right\rfloor, \qquad v_{t,n} = \left\lfloor \frac{y_{t,n} - y_{\min}}{y_{\max} - y_{\min}} H \right\rfloor. \tag{34}$$

**Table 5: Complete LUMI task-model benchmark. Bold entries indicate the strongest result within each workload and metric.**

| Workload | Method | mAP | mIoU | Acc@.50 |
|---|---|---|---|---|
| Simulation | PillarNet | 0.47 | 0.66 | 0.74 |
| | PointNet++ | 0.49 | 0.70 | 0.81 |
| | BEV-MAE | 0.42 | 0.79 | 0.87 |
| | GroupFree3D | 0.53 | 0.68 | 0.79 |
| | **AFN** | **0.90** | **0.86** | **0.97** |
| Real | PillarNet | 0.33 | **0.81** | 0.89 |
| | PointNet++ | 0.34 | 0.80 | 0.90 |
| | BEV-MAE | 0.30 | 0.69 | 0.79 |
| | GroupFree3D | 0.34 | 0.76 | 0.85 |
| | **AFN** | **0.63** | 0.72 | **0.91** |
| LiDAR-Net | PillarNet | 0.23 | 0.55 | 0.50 |
| | PointNet++ | 0.21 | 0.58 | 0.55 |
| | BEV-MAE | 0.19 | 0.52 | 0.45 |
| | GroupFree3D | 0.24 | 0.57 | 0.52 |
| | **AFN** | **0.29** | **0.62** | **0.60** |

For each occupied cell $(u, v)$, the six-channel BEV representation stores

$$\mathbf{X}_{t,:,u,v} = \left[\mu^z_{t,u,v}, z^{\min}_{t,u,v}, z^{\max}_{t,u,v}, \sigma^z_{t,u,v}, \mu^r_{t,u,v}, \rho_{t,u,v}\right], \tag{35}$$

where

$$\rho_{t,u,v} = \min\left(\frac{N_{t,u,v}}{10}, 1\right). \tag{36}$$

Empty cells receive zero-valued features. When intensity is disabled, the intensity channel is zero. The aligned coordinate tensor is

$$\mathbf{C}_t = [\mathbf{C}^x, \mathbf{C}^y, \mathbf{C}^z_t] \in \mathbb{R}^{H\times W\times 3}, \tag{37}$$

where the planar coordinate fields are fixed by the BEV grid and

$$\mathbf{C}^z_{t,u,v} = \mu^z_{t,u,v} \tag{38}$$

for representations that include mean height. Density-only ablations set the third coordinate channel to zero to prevent hidden mean-height leakage.

## C.2 Weighted Height Encoding

AFN uses $L = 6$ frequency levels,

$$\omega_\ell = 2^\ell \pi, \qquad \ell = 0, \ldots, L-1. \tag{39}$$

For scalar height $z$, the unweighted encoding is

$$\mathbf{g}(z) = [\sin(\omega_0 z), \ldots, \sin(\omega_{L-1} z), \cos(\omega_0 z), \ldots, \cos(\omega_{L-1} z)]. \tag{40}$$

The implementation applies

$$w_j = \exp(-j/2), \qquad j = 0, \ldots, 2L-1, \tag{41}$$

giving

$$\boldsymbol{\gamma}(z) = \mathbf{w} \odot \mathbf{g}(z). \tag{42}$$

The complete AFN input is

$$\mathbf{F}^0_t = \mathrm{Concat}[\mathbf{X}_t, \boldsymbol{\gamma}(\mathbf{C}^z_t)] \in \mathbb{R}^{18\times H\times W}. \tag{43}$$

The weights are applied after concatenating all sine channels followed by all cosine channels, matching the Quality implementation.

## C.3 Coordinate-Conditioned Perspective Branches

Each perspective branch applies an independent transform

$$\overline{\mathbf{F}}^b_t = \phi_b(\mathbf{F}^0_t), \qquad b \in \{\mathrm{xy}, \mathrm{xz}, \mathrm{yz}\}, \tag{44}$$

where

$$\phi_b = \mathrm{ReLU} \circ \mathrm{BN} \circ \mathrm{Conv}^b_{3\times 3}. \tag{45}$$

The coordinate modulations are

$$\Gamma_{\mathrm{xy}} = \sin(\alpha_{\mathrm{xy}} \mathbf{C}^z_t), \tag{46}$$

$$\Gamma_{\mathrm{xz}} = \sin(\alpha_{\mathrm{xz}} \mathbf{C}^z_t) + \cos(\alpha_{\mathrm{xz}} \mathbf{C}^x), \tag{47}$$

$$\Gamma_{\mathrm{yz}} = \sin(\alpha_{\mathrm{yz}} \mathbf{C}^z_t) + \cos(\alpha_{\mathrm{yz}} \mathbf{C}^y), \tag{48}$$

and

$$\mathbf{F}^b_t = \overline{\mathbf{F}}^b_t \odot (1 + \Gamma_b). \tag{49}$$

The three convolutions and modulation scalars are independently parameterized.

## C.4 Shared Multi-Dilation Denoising

For active dilation set $\mathcal{D}_p$, the denoising module is

$$\mathcal{M}_{\mathcal{D}_p}(\mathbf{F}) = \mathbf{F} + \mathrm{Conv}^p_{1\times 1}\left(\Big\Vert_{d\in\mathcal{D}_p} \mathcal{D}^p_d(\mathbf{F})\right), \tag{50}$$

where $\mathcal{D}^p_d$ is a 3×3 convolution with dilation $d$. The module instance is reused across perspective responses:

$$\widehat{\mathbf{F}}^b_t = \mathcal{M}_{\mathcal{D}_p}(\mathbf{F}^b_t). \tag{51}$$

Quality uses

$$\mathcal{D}_{\mathrm{Q}} = \{1, 2, 4\}, \tag{52}$$

whereas D14 uses

$$\mathcal{D}_{\mathrm{D14}} = \{1, 4\}. \tag{53}$$

Quality and D14 have separately trained $1 \times 1$ fusion layers. D14 is not constructed by deleting trained Quality weights.

## C.5 Adaptive Geometric-Axis Fusion

The Quality implementation computes scalar variance over each complete denoised branch tensor,

$$v_b = \mathrm{Var}(\widehat{\mathbf{F}}^b_t). \tag{54}$$

Given learned coefficients $a_b$, the normalized coefficients are

$$\beta_b = \frac{\exp(-v_b) a_b}{\max\left(\sum_j \exp(-v_j) a_j, \epsilon\right)}. \tag{55}$$

Let

$$c(i, j) = \frac{1}{B} \sum_{n=1}^{B} \mathrm{CosSim}(\mathrm{vec}\,\widehat{\mathbf{F}}^i_{t,n}, \mathrm{vec}\,\widehat{\mathbf{F}}^j_{t,n}) \tag{56}$$

denote minibatch-averaged cosine similarity. The global consistency coefficient is

$$\kappa_t = \frac{c(\mathrm{xz}, \mathrm{yz}) + c(\mathrm{xz}, \mathrm{xy}) + c(\mathrm{yz}, \mathrm{xy})}{3}, \tag{57}$$

and

$$\mathbf{F}^{\mathrm{enc}}_t = \kappa_t \,\mathrm{BN}\left(\sum_b \beta_b \widehat{\mathbf{F}}^b_t\right). \tag{58}$$

A hardware realization of this fusion operator must preserve the distinction between externally supplied gated accumulation and the complete Quality operator. The latter additionally computes branch variances, normalized coefficients, the cosine-consistency term, and BatchNorm. Hardware realization details are deferred to a later unified hardware appendix.

## C.6 Global-Local Context Backbone

A lightweight encoder-decoder produces

$$\mathbf{F}_t^{\text{local}} = \text{FPN}(\mathbf{F}_t^{\text{enc}}). \tag{59}$$

Global tokens are generated from pooled encoder features,

$$\mathbf{p}_t = \text{AvgPool}(\mathbf{F}_t^{\text{enc}}), \tag{60}$$

$$\mathbf{T}_{t,i} = \mathbf{E}_i + \mathbf{W}_p \mathbf{p}_t, \qquad i = 1, \ldots, N_p. \tag{61}$$

The projected token context is

$$\mathbf{c}_t = \frac{1}{N_p} \sum_{i=1}^{N_p} \mathbf{W}_t \mathbf{T}_{t,i}, \tag{62}$$

and is broadcast spatially. The local and global representations are combined through $1 \times 1$ projections, squeeze-and-excitation, batch normalization, and ReLU.

## C.7 Dense Targets and Regression Encoding

For an object with physical length $l$ and width $w$, the corresponding grid dimensions are

$$l_g = \frac{l}{x_{\max} - x_{\min}} W, \qquad w_g = \frac{w}{y_{\max} - y_{\min}} H. \tag{63}$$

The target-adaptive Gaussian scale and support radius are

$$\sigma = \max\left(\frac{\max(l_g, w_g)}{6}, 1\right), \tag{64}$$

$$r = \left\lfloor \frac{\max(l_g, w_g)}{2} \right\rfloor + 2. \tag{65}$$

For continuous grid center $(g_x^{\text{float}}, g_y^{\text{float}})$, the implementation uses integer center

$$g_x = \lfloor g_x^{\text{float}} \rfloor, \qquad g_y = \lfloor g_y^{\text{float}} \rfloor. \tag{66}$$

The Gaussian response is

$$Q_{x,y} = \exp\left(-\frac{(x - g_x)^2 + (y - g_y)^2}{2\sigma^2}\right). \tag{67}$$

A cell belongs to the target support iff

$$(x - g_x)^2 + (y - g_y)^2 < r^2 \quad \text{and} \quad Q_{x,y} > 0.1. \tag{68}$$

The formal runs use annotation-order overlap assignment. A later annotation in the label file overwrites an earlier target in shared support cells.

For every positive support cell $(x, y)$, the regression target is

$$\mathbf{y}_{x,y}^{\text{reg}} = \left(g_x^{\text{float}} - x, g_y^{\text{float}} - y, \log(l_g + 10^{-6}), \log(w_g + 10^{-6}), \cos 2\theta, \sin 2\theta\right). \tag{69}$$

The regression mask is active exactly on the assigned support.

**Table 6: AFN training and checkpoint-selection protocol.**

| Setting | Simulation | Real |
|---|---|---|
| Input resolution | 400 × 400 | |
| Input channels | 6 | |
| Foreground classes | 5 | 4 |
| Representation | Full | |
| Feature width | 64 | |
| Height levels | 6 | |
| Orientation | Double angle | |
| Regression channels | 6 | |
| Train/val./test | 2,040/255/255 | 416/52/52 |
| Point sampling | Disabled | |
| Optimizer | Adam | |
| Batch size | 4 | |
| Max. epochs | 500 | |
| Learning rate | $10^{-4}$ | |
| Weight decay | $10^{-4}$ | |
| LR schedule | StepLR, 30 epochs, $\gamma = 0.1$ | |
| Classification loss | Focal, $\alpha = 2$, $\gamma = 4$ | |
| Box loss | Smooth L1 | |
| Regression weight | 1.0 | |
| Yaw weight | 1.0 | |
| Density norm. | 10.0 | |
| Target threshold | 0.1 | |
| Loader workers | 6 | |
| Checkpoint criterion | Min. validation loss | |
| Checkpoint interval | 25 epochs | |
| Early-stop warm-up | 30 epochs | |
| Early-stop patience | 45 epochs | |
| Early-stop $\Delta_{\min}$ | $10^{-4}$ | |
| Random seed | 42 | |

## C.8 Loss, Optimization, and Checkpoint Selection

The classification objective is categorical focal loss with $\alpha = 2$ and $\gamma = 4$. The non-yaw box terms use Smooth L1 loss. For double-angle orientation, the dedicated double-angle objective combines box and yaw terms. The complete loss is

$$\mathcal{L} = \mathcal{L}_{\text{cls}} + \lambda_{\text{reg}}(\mathcal{L}_{\text{box}} + \lambda_{\text{yaw}} \mathcal{L}_{\text{yaw}}), \tag{70}$$

with $\lambda_{\text{reg}} = \lambda_{\text{yaw}} = 1$.

Simulation and Real use Adam, batch size four, initial learning rate $10^{-4}$, weight decay $10^{-4}$, and at most 500 epochs. StepLR multiplies the learning rate by 0.1 every 30 epochs. The best checkpoint is selected by minimum validation total loss. Improvement requires a reduction of at least $10^{-4}$. Early stopping is enabled after a 30-epoch warm-up and triggers after 45 consecutive non-improving epochs. The latest checkpoint is saved after every epoch, the best checkpoint whenever validation loss improves, and periodic checkpoints every 25 epochs.

# D Profile Space and Conditional Interaction Analysis

## D.1 Profile Representation

An AFN profile is represented as

$$p = (N_p, C_p^{\text{head}}, \mathcal{D}_p, \mathcal{A}_p, C_p^{\text{feat}}), \tag{71}$$

where the entries denote global-token count, prediction-head hidden width, active dilation set, geometric-axis fusion mechanism, and common feature width. The Quality reference is

$$p_{\text{Q}} = (4, 32, \{1, 2, 4\}, \mathcal{A}_{\text{geo}}, 64). \tag{72}$$

## D.2 Signed Normalized Metric Change

For workload $d$, transformation $a$, and matched Quality reference $r$, the signed normalized change is

$$\Delta_j(a; r, d) = \delta_j \frac{q_j(a, d) - q_j(r, d)}{\max(|q_j(r, d)|, \epsilon)}, \tag{73}$$

where $\delta_j = +1$ for larger-is-better metrics and $-1$ for smaller-is-better metrics. Positive values always denote improvement.

## D.3 Pairwise Transformation Interaction

For directly trained transformations $a$ and $b$,

$$I_j(a, b; r, d) = \Delta_j(a \oplus b; r, d) - \Delta_j(a; r, d) - \Delta_j(b; r, d). \tag{74}$$

A negative value indicates that the measured composition underperforms the additive expectation from isolated transformations. The statistic is descriptive and is not evaluated for untrained combinations.

## D.4 Higher-Order Conditional Increment

For transformation $a$ added to an existing composition $c$,

$$\Delta_j(a \mid c; d) = \delta_j \frac{q_j(a \oplus c, d) - q_j(c, d)}{\max(|q_j(c, d)|, \epsilon)}. \tag{75}$$

# E Detailed FPGA Methodology

This appendix expands the FPGA methodology introduced in Section 4. It describes how FieldRT converts a geometry-preserving AFN profile into a parameterized programmable-logic architecture, including contract representation, profile resolution, spatial lowering, compile-time specialization, runtime control, stream composition, memory organization, and numerical qualification.

## E.1 FPGA Methodology Overview

FieldRT treats hardware realization as a continuation of geometric profile construction. Each qualified AFN profile defines a set of retained geometric operations, tensor edges, numerical formats, consumer relationships, and state lifetimes. FieldRT records these properties in a geometric execution contract and resolves the contract into a hardware configuration.

The FPGA methodology follows five steps:

(1) construct a qualified AFN profile from instance-level geometric sensitivity;
(2) describe the retained fields and tensor edges through geometric execution descriptors;
(3) resolve the descriptors into topology, dataflow, numerical, storage, and control parameters;
(4) lower recurring geometric structures into reusable spatial dataflow objects; and
(5) qualify each generated operation against an operation-matched reference before admitting the configuration into profile evaluation.

This workflow preserves a direct relationship between a profile decision and its hardware consequences. Changes in fusion, dilation, feature capacity, prediction width, and global context alter both model behavior and the corresponding dataflow, storage, and control configuration.

## E.2 Geometric Execution Contract

For an AFN profile $p$, FieldRT defines the geometric execution contract

$$\Gamma(p) = [\mathcal{F}(p), \mathcal{E}(p), \mathcal{N}(p), \mathcal{T}(p), \mathcal{L}(p)], \tag{76}$$

where:

- $\mathcal{F}$ describes geometric-field semantics, tensor shape, spatial extent, and coordinate convention;
- $\mathcal{E}$ describes producer–consumer edges, consumer multiplicity, and reuse;
- $\mathcal{N}$ specifies storage formats, accumulation formats, scaling conventions, and numerical requirements;
- $\mathcal{T}$ describes traversal order, branch topology, reduction structure, and materialization boundaries; and
- $\mathcal{L}$ describes the lifetime of field values, neighborhoods, partial sums, and persistent state.

The contract separates geometric meaning from an incidental operator sequence. Two tensors with identical shapes can receive different execution descriptors when their geometric roles, consumers, numerical constraints, or lifetimes differ.

*E.2.1 Profile descriptor.* A profile-level descriptor records the major structural choices:

$$\boldsymbol{\pi}(p) = \left[N_p, C_{\text{head}}, C_{\text{feat}}, \mathcal{D}, A, G\right], \tag{77}$$

where $N_p$ is global-token capacity, $C_{\text{head}}$ is prediction-head width, $C_{\text{feat}}$ is field-feature width, $\mathcal{D}$ is the retained dilation set, $A$ describes axis-fusion behavior, and $G$ identifies the supported numerical configuration.

These variables have direct hardware interpretations:

- $N_p$ determines token and reduction-state capacity;
- $C_{\text{head}}$ determines prediction traversal and head arithmetic dimensions;
- $C_{\text{feat}}$ determines channel capacity, register width, and parallel-lane organization;
- $\mathcal{D}$ determines active dilation branches, selectable taps, and line-buffer depth;
- $A$ determines gate granularity, gate storage, and fusion control; and
- $G$ determines field storage, weight, accumulator, and output formats.

**Table 7: Controlled AFN candidate matrix. Retained profiles undergo formal TensorRT deployment, while sensitivity and interaction profiles support controlled algorithmic analysis.**

| Profile | Tokens | Head | Dilations | Fusion | Width | Role | Workload |
|---|---|---|---|---|---|---|---|
| Quality | 4 | 32 | $\{1,2,4\}$ | Geo. | 64 | Reference | Sim., Real |
| Tokens-2 | 2 | 32 | $\{1,2,4\}$ | Geo. | 64 | Retained, isolated | Sim., Real |
| Head-16 | 4 | 16 | $\{1,2,4\}$ | Geo. | 64 | Retained, isolated | Sim., Real |
| D14 | 4 | 32 | $\{1,4\}$ | Geo. | 64 | Retained, isolated | Sim., Real |
| H16-D14 | 4 | 16 | $\{1,4\}$ | Geo. | 64 | Retained, combined | Sim., Real |
| Feat-48 | 4 | 24 | $\{1,2,4\}$ | Geo. | 48 | Width sensitivity | Sim. |
| G2 | 4 | 32 | $\{1,2,4\}$ | Static channel | 64 | Fusion sensitivity | Sim. |
| G3a | 4 | 32 | $\{1,2,4\}$ | Sample scalar | 64 | Fusion sensitivity | Sim. |
| G3b | 4 | 32 | $\{1,2,4\}$ | Sample channel | 64 | Fusion sensitivity | Sim. |
| T2-H16 | 2 | 16 | $\{1,2,4\}$ | Geo. | 64 | Pair interaction | Sim. |
| T2-D14 | 2 | 32 | $\{1,4\}$ | Geo. | 64 | Pair interaction | Sim. |
| T2-H16-D14 | 2 | 16 | $\{1,4\}$ | Geo. | 64 | Higher-order interaction | Sim. |

*E.2.2 Edge descriptor.* For a tensor edge $e = (u, v)$, FieldRT records

$$\gamma_e = [\mathbf{s}_e, \nu_e, \rho_e, \lambda_e, \eta_e, \mu_e], \tag{78}$$

where $\mathbf{s}_e$ is the tensor shape, $\nu_e$ is the numerical representation, $\rho_e$ describes consumer multiplicity and reuse, $\lambda_e$ is the state lifetime, $\eta_e$ records geometric sensitivity, and $\mu_e$ specifies the materialization policy.

The materialization policy determines whether an edge remains in an AXI4-Stream path, enters local storage, or crosses an external-memory boundary. Consumer multiplicity informs broadcast or replication, while lifetime determines the amount of retained state.

## E.3 Contract Resolution

The Contract Fabric resolves the profile and edge descriptors into a backend-specific hardware configuration:

$$\boldsymbol{\theta}_{\mathrm{hw}} = \mathcal{R}_{\mathrm{PL}}\left(\Gamma(p)\right) = \left[\boldsymbol{\theta}_{\mathrm{top}}, \boldsymbol{\theta}_{\mathrm{flow}}, \boldsymbol{\theta}_{\mathrm{num}}, \boldsymbol{\theta}_{\mathrm{buf}}, \boldsymbol{\theta}_{\mathrm{ctrl}}\right]. \tag{79}$$

The five resolved parameter groups represent:

- $\boldsymbol{\theta}_{\mathrm{top}}$: active kernels, branches, producers, consumers, and stream connections;
- $\boldsymbol{\theta}_{\mathrm{flow}}$: traversal order, channel parallelism, reduction structure, and materialization boundaries;
- $\boldsymbol{\theta}_{\mathrm{num}}$: input, weight, accumulation, activation, and output formats;
- $\boldsymbol{\theta}_{\mathrm{buf}}$: local storage dimensions, line-buffer lifetime, register depth, and FIFO capacity; and
- $\boldsymbol{\theta}_{\mathrm{ctrl}}$: mode fields, loop bounds, masks, address bases, transfer phases, and completion conditions.

Resolution proceeds from topology to storage and control. The topology resolver determines active consumers and branch connections. The dataflow resolver chooses streaming and materialized boundaries and assigns local state. The numerical resolver determines compatible storage and accumulation formats. The control resolver generates the register and sequencing configuration required by the resolved datapath.

## E.4 Geometry-Native Spatial Objects

FieldRT maps recurring geometric relationships to reusable spatial dataflow objects. The principal objects are single-read field distribution, coordinate-conditioned processing, streaming geometric fusion, lifetime-aware multi-dilation buffering, streaming global aggregation, and local–global prediction.

*E.4.1 Single-read field distribution.* AFN applies complementary axis-conditioned transformations to a common BEV field. The Multi-Field Engine traverses the field once and distributes aligned activation values to the $xy$, $xz$, and $yz$ consumers:

$$\mathbf{X}_t \longrightarrow \left\{\mathbf{X}_t^{xy}, \mathbf{X}_t^{xz}, \mathbf{X}_t^{yz}\right\}. \tag{80}$$

For $B$ consumers, branch-local field reads require

$$A_{\mathrm{field}}^{\mathrm{branch}} = B|\mathbf{X}|, \tag{81}$$

whereas single-read distribution requires

$$A_{\mathrm{field}}^{\mathrm{shared}} = |\mathbf{X}|. \tag{82}$$

The fan-out shares activation access while preserving independent coordinate and coefficient streams for each geometric branch. Branch outputs carry a common traversal index so that downstream fusion receives aligned axis-conditioned responses.

*E.4.2 Coordinate-conditioned processing.* Each branch combines an activation value with an axis-conditioned coordinate and branch-specific coefficient. A representative processing element computes

$$\mathbf{z}_{t,a} = \mathcal{A}\left(\mathbf{x}_t \odot \mathbf{c}_{t,a} \odot \mathbf{w}_a + \mathbf{b}_a\right), \qquad a \in \{xy, xz, yz\}, \tag{83}$$

where $\mathbf{c}_{t,a}$ is the coordinate representation associated with axis $a$, $\mathbf{w}_a$ and $\mathbf{b}_a$ are branch parameters, and $\mathcal{A}$ denotes the selected accumulation and activation rule.

Separate activation, coordinate, and weight registers align the three input classes before wide accumulation. Coordinate storage follows the resolved reuse descriptor, permitting recurrent coordinate values to remain locally available across field traversal.

#### E.4.3 Streaming geometric fusion.
The fusion object combines aligned axis-conditioned streams according to

$$\mathbf{S}_t = \sum_{a \in \{xy, xz, yz\}} \boldsymbol{\alpha}_{t,a} \odot \mathbf{Z}_{t,a}. \tag{84}$$

The gate $\boldsymbol{\alpha}_{t,a}$ supports four interface modes:

(1) static scalar;
(2) static channel-wise;
(3) streamed scalar; and
(4) streamed channel-wise.

The gate mode determines the size and update behavior of the gate registers. Multipliers apply branch weights, and an adder tree produces one fused output stream. Fusion shares a producer–consumer boundary with the branch path, so the branch streams can be consumed in traversal order.

#### E.4.4 Lifetime-aware multi-dilation buffering.
For active dilation set $\mathcal{D}(p)$, a $3 \times 3$ branch with dilation $d$ accesses

$$\mathcal{W}_d(i,j) = \{X(i+rd, j+sd) \mid r,s \in \{-1,0,1\}\}. \tag{85}$$

The maximum active dilation determines the retained row lifetime:

$$R_{\mathrm{buf}}(p) = 2 \max_{d \in \mathcal{D}(p)} d + 1. \tag{86}$$

FieldRT implements this lifetime with one physical line buffer and shift register state. A window generator constructs the current maximum-lifetime window, and a selectable tap network exposes branch-specific neighborhoods. For coordinate pair $(r,s)$, branch $d$ selects the offset $(rd, sd)$.

The active branches share the line-buffered activation field. Each branch retains its own coefficient selection, partial-sum state, output quantization, and output stream. Adding a retained dilation branch consequently alters tap and accumulation topology while leaving the activation-storage lifetime bounded by $D_{\max}$.

#### E.4.5 Streaming global aggregation.
For a field $\mathbf{X} \in \mathbb{R}^{C \times H \times W}$, FieldRT represents global aggregation as persistent channel state:

$$g_c^{(t+1)} = g_c^{(t)} + x_{c,t}, \qquad c = 1, \ldots, C. \tag{87}$$

After the final spatial element, normalization and projection generate the global representation used by the token path. The storage requirement follows channel and token capacity:

$$M_{\mathrm{global}} = O(C + N_p C_{\mathrm{head}}), \tag{88}$$

rather than the complete spatial field. The token representation enters local–global prediction through an explicit stream or local-buffer boundary.

#### E.4.6 Local–global prediction.
The local prediction stream and the global token stream are synchronized at the local–global merge. The merged representation is consumed by the classification and regression heads using a shared traversal interface:

$$\mathbf{h}_t = \mathcal{M}\left(\mathbf{f}_t^{\mathrm{local}}, \mathbf{g}\right), \qquad \left(\hat{\mathbf{y}}_t^{\mathrm{cls}}, \hat{\mathbf{y}}_t^{\mathrm{reg}}\right) = \mathcal{H}(\mathbf{h}_t). \tag{89}$$

The profile controls the local feature width, token capacity, and head width. These variables determine merge dimensions, token-register capacity, and head traversal bounds in the resolved architecture.

## E.5 Compile-Time Specialization

The generated datapath uses compile-time parameters to establish its maximum supported envelope:

$$\boldsymbol{\theta}_{\mathrm{compile}} = \left[C_{\max}, H, W, D_{\max}, B_{\max}, P_{\mathrm{ch}}, W_{\mathrm{in}}, W_{\mathrm{w}}, W_{\mathrm{acc}}, \mathcal{M}_{\mathrm{gate}}\right], \tag{90}$$

where $B_{\max}$ is the maximum branch count, $P_{\mathrm{ch}}$ is channel parallelism, $W_{\mathrm{in}}$ and $W_{\mathrm{w}}$ are activation and coefficient widths, $W_{\mathrm{acc}}$ is accumulation width, and $\mathcal{M}_{\mathrm{gate}}$ is the supported fusion-mode set.

Compile-time parameters determine:

- line-buffer dimensions and memory banking;
- the maximum number of branch and channel lanes;
- register and accumulator widths;
- FIFO data widths and local capacities;
- the available tap and gate modes; and
- the AXI4-Stream and AXI-Lite interface structure.

The compiled envelope bounds storage and arithmetic resources before execution. Profiles whose resolved requirements fit this envelope can share the same parameterized datapath.

## E.6 Runtime Configuration and Control

Within the compiled envelope, a runtime configuration selects an active subconfiguration:

$$\boldsymbol{\theta}_{\mathrm{run}} = \left[m_{\mathrm{profile}}, \mathbf{m}_{\mathrm{branch}}, \mathbf{m}_{\mathrm{lane}}, m_{\mathrm{gate}}, m_{\mathrm{precision}}, \mathbf{a}_{\mathrm{base}}, \mathbf{n}_{\mathrm{transfer}}\right]. \tag{91}$$

Here, $m_{\mathrm{profile}}$ identifies the resolved profile, $\mathbf{m}_{\mathrm{branch}}$ and $\mathbf{m}_{\mathrm{lane}}$ activate branches and channel lanes, $m_{\mathrm{gate}}$ selects fusion behavior, $m_{\mathrm{precision}}$ selects the supported numerical mode, $\mathbf{a}_{\mathrm{base}}$ contains external-memory address bases, and $\mathbf{n}_{\mathrm{transfer}}$ contains logical transfer lengths.

The control plane contains four principal components.

*Profile and mode registers.* The profile register file stores active bounds, masks, supported numerical formats, and mode identifiers. Software writes one coherent configuration before launching a transfer phase.

*Configuration decoder.* The decoder converts profile fields into branch enables, lane masks, bypass signals, tap selections, gate-register behavior, and output routing.

*Loop and address control.* Nested counters generate spatial and channel traversal. Address controllers produce local-memory indices and DMA descriptor progression from the resolved tensor shape and layout.

*Status counters.* Status registers return cycle counts, transferred bytes, occupancy, completion state, and numerical flags. The Contract Fabric associates these measurements with the profile and execution configuration that produced them.

## E.7 AXI4-Stream Composition

The programmable-logic datapath represents tensor movement through AXI4-Stream channels. Each stream carries data together with a logical element order and a tensor-completion contract. Elastic FIFOs decouple adjacent stages, while transfer-phase control aligns kernels that consume different input classes.

For a stream connection from producer $u$ to consumer $v$, FieldRT records

$$\sigma_{u\to v} = \left[W_{\text{data}}, N_{\text{logical}}, \tau_{\text{last}}, d_{\text{fifo}}, \phi_{\text{transfer}}\right], \tag{92}$$

where $W_{\text{data}}$ is stream width, $N_{\text{logical}}$ is the logical tensor length, $\tau_{\text{last}}$ defines tensor completion, $d_{\text{fifo}}$ is elastic depth, and $\phi_{\text{transfer}}$ is the active transfer phase.

This descriptor allows the Contract Fabric to connect kernels whose internal arithmetic and traversal differ while preserving one explicit tensor boundary.

## E.8 Phased Input Routing

Multi-input kernels receive coefficients, gates, coordinates, and activations through explicit transfer phases. The control tuple for phase $q$ is

$$\phi_q = \left[r_q, N_q, a_q, s_q\right], \tag{93}$$

where $r_q$ selects the destination route, $N_q$ gives the logical beat count, $a_q$ gives the source address, and $s_q$ launches the transfer.

The phase controller commits the route and length before asserting start. Completion returns the route to idle before the next phase is configured. This sequence provides deterministic association between a software transfer and the receiving kernel interface.

For shared-buffer denoising, the resulting path is

$$\begin{aligned}\text{MM2S} &\longrightarrow \text{controlled demultiplexer} \longrightarrow \text{coefficient/activation ports}\\ &\longrightarrow \text{shared line buffer} \longrightarrow \text{branch accumulators} \longrightarrow \text{S2MM}.\end{aligned} \tag{94}$$

## E.9 Logical Tensor Boundaries Across DMA Descriptors

A logical tensor can span several DMA descriptors. Let descriptor $j$ contain $n_j$ stream beats. The complete logical transfer contains

$$N_{\text{tensor}} = \sum_{j=1}^{J} n_j. \tag{95}$$

FieldRT programs $N_{\text{tensor}}$ into the receiving route controller. The controller increments a logical beat counter for every accepted stream element and emits the tensor-completion event when

$$n_{\text{accepted}} = N_{\text{tensor}}. \tag{96}$$

This mechanism separates the tensor boundary from the segmentation used by the memory-transfer engine. Kernel traversal, window state, branch accumulation, and output completion therefore follow the logical tensor description in the execution contract.

## E.10 Elastic Output Drainage

Each materialized branch output is connected to an elastic FIFO and a dedicated S2MM path. The FIFO absorbs short-term readiness differences between the compute pipeline and memory transfer. The S2MM channel is armed before the corresponding producer starts, allowing accepted outputs to drain continuously.

For branch set $\mathcal{B}_{\text{out}}$, the output arrangement is

$$\forall b \in \mathcal{B}_{\text{out}}: \qquad \text{producer}_b \longrightarrow \text{FIFO}_b \longrightarrow \text{S2MM}_b. \tag{97}$$

Independent output paths preserve branch ordering and completion state. If the resolved contract fuses the branches on chip, only the fused stream receives an external output path.

## E.11 Numerical Representation

The numerical descriptor specifies the format of each storage and arithmetic boundary:

$$\nu_e = [W_e, S_e, F_e, Z_e, \mathcal{R}_e, \mathcal{S}_e], \tag{98}$$

where $W_e$ is total width, $S_e$ identifies signedness, $F_e$ is fractional precision or fixed-point position, $Z_e$ is zero-point information where applicable, $\mathcal{R}_e$ is the rounding rule, and $\mathcal{S}_e$ is the saturation rule.

For a multiply–accumulate path, the resolver selects an accumulator width satisfying

$$W_{\text{acc}} \geq W_{\text{in}} + W_{\text{w}} + \lceil \log_2 K_{\text{sum}} \rceil, \tag{99}$$

where $K_{\text{sum}}$ is the maximum number of accumulated terms. The resolved output stage applies the profile-specific shift, rounding, saturation, and activation convention.

Numerical modes are part of the compiled contract. A mode selection changes only among formats and arithmetic behaviors represented by the generated datapath.

## E.12 Operation-Matched Reference Construction

For every custom-PL arithmetic object, FieldRT constructs a software reference from the resolved execution contract. The reference uses the same:

- tensor shape and traversal order;
- branch and tap selection;
- coefficient layout;
- integer width and signedness;
- accumulation order;
- shift, rounding, saturation, and activation behavior; and
- output layout and materialization boundary.

For hardware function $H_\Gamma$ and reference function $R_\Gamma$ generated from contract $\Gamma$, qualification evaluates

$$\mathbf{y}^{\text{hw}} = H_\Gamma(\mathbf{x}, \mathbf{w}, \boldsymbol{\theta}_{\text{run}}), \qquad \mathbf{y}^{\text{ref}} = R_\Gamma(\mathbf{x}, \mathbf{w}, \boldsymbol{\theta}_{\text{run}}). \tag{100}$$

The comparison reports

$$N_{\neq} = \sum_{k=1}^{K} \mathbf{1}\left[y_k^{\text{hw}} \neq y_k^{\text{ref}}\right], \qquad \text{MAE} = \frac{1}{K}\sum_{k=1}^{K}\left|y_k^{\text{hw}} - y_k^{\text{ref}}\right|. \tag{101}$$

Integer outputs satisfy the exact qualification rule when

$$N_{\neq} = 0 \quad \wedge \quad \mathrm{MAE} = 0. \tag{102}$$

Floating-point outputs additionally report finite-output status, maximum absolute error, and cosine similarity.

## E.13 Three-Stage Qualification Workflow

FieldRT applies the same reference and fixture through three execution stages:

(1) operation-matched software execution;
(2) HLS C simulation; and
(3) physical programmable-logic execution.

The software-to-C-simulation comparison validates the HLS operation before synthesis. The C-simulation-to-board comparison validates the synthesized interface, control sequence, stream composition, and numerical datapath. The software-to-board comparison closes the complete operation-level path.

Each stage records the contract identifier, input and coefficient hashes, tensor dimensions, numerical descriptor, active modes, and output hash. These identifiers bind all qualification stages to the same operation instance.

## E.14 Hardware Feedback to the Contract Fabric

A resolved profile defines the measurements returned by the hardware path:

$$\mathbf{r}_{\mathrm{hw}} = \left[\ell_{\mathrm{txn}}, B_{\mathrm{in}}, B_{\mathrm{out}}, O_{\mathrm{fifo}}, U_{\mathrm{LUT}}, U_{\mathrm{FF}}, U_{\mathrm{DSP}}, U_{\mathrm{BRAM}}, s_{\mathrm{num}}\right], \tag{103}$$

where $\ell_{\mathrm{txn}}$ is transaction latency, $B_{\mathrm{in}}$ and $B_{\mathrm{out}}$ are materialized input and output traffic, $O_{\mathrm{fifo}}$ represents stream occupancy, $U_{\cdot}$ gives resource utilization, and $s_{\mathrm{num}}$ records numerical status.

The Contract Fabric combines these observations into profile cost and eligibility:

$$\left(C_{\mathrm{sys}}(p, \mathrm{PL}), \chi_{\mathrm{eligible}}(p, \mathrm{PL})\right) = \mathcal{A}\left(\mathbf{r}_{\mathrm{hw}}, \mathbf{Q}(p, \mathrm{PL})\right), \tag{104}$$

where $\mathcal{A}$ is the cost-and-validity aggregation function. This feedback closes the loop from profile construction to contract resolution, physical execution, and profile eligibility.

# F Complete Simulation and Real Profile Results

## F.1 Simulation PyTorch Results

Table 8 reports the complete Simulation-side PyTorch results for the reference profile, the retained deployment candidates, and the additional structural controls. These results provide the task-quality baseline for the subsequent TensorRT regression and deployment analysis.

## F.2 Real PyTorch Results

Table 9 reports the verified Real-side PyTorch results for the reference profile and the two evaluated compact profiles. These results characterize the cross-workload sensitivity of the retained transformations and provide the task-quality reference for the corresponding TensorRT evaluation.

**Table 8: Simulation PyTorch results for the evaluated AFN profiles.**

| Profile | mAP | $AP_{50}$ | $AP_{75}$ | mIoU | Dense mAP |
|---|---|---|---|---|---|
| Quality | 0.596 | 0.921 | 0.684 | 0.779 | 0.853 |
| Tokens-2 | 0.604 | 0.933 | 0.684 | 0.777 | 0.845 |
| Head-16 | 0.601 | 0.937 | 0.697 | 0.777 | 0.843 |
| D14 | 0.601 | 0.940 | 0.678 | 0.782 | 0.861 |
| H16-D14 | 0.598 | 0.923 | 0.662 | 0.779 | 0.840 |
| Feat-48 | 0.516 | 0.886 | 0.524 | 0.748 | 0.785 |
| G2 | 0.548 | 0.916 | 0.589 | 0.765 | 0.834 |
| G3a | 0.578 | 0.925 | 0.634 | 0.767 | 0.833 |
| G3b | 0.547 | 0.912 | 0.605 | 0.759 | 0.820 |

**Table 9: Real PyTorch results for the evaluated AFN profiles.**

| Profile | mAP | $AP_{50}$ | $AP_{75}$ | mIoU | Dense mAP |
|---|---|---|---|---|---|
| Quality | 0.106 | 0.395 | 0.013 | 0.532 | 0.460 |
| Tokens-2 | 0.070 | 0.287 | 0.010 | 0.481 | 0.490 |
| H16-D14 | 0.057 | 0.244 | 0.005 | 0.467 | 0.413 |

## F.3 TensorRT FP32 Complete-Test Results

Table 10 reports the complete TensorRT FP32 task results for all retained profile–workload engines. The comparison covers instance detection, strict localization, dense prediction, and the full evaluation-set size for both Simulation and Real.

# G TensorRT Artifact and Qualification Protocol

## G.1 Audited Artifact Pipeline

Each profile follows

$$\text{Checkpoint} \rightarrow \text{Strict Model Load} \rightarrow \text{ONNX} \rightarrow \text{TensorRT Engine} \rightarrow \text{Deploymen} \tag{105}$$

Checkpoint keys may undergo only documented data-parallel prefix removal or an explicitly recorded legacy-key migration. After normalization,

$$\mathcal{K}_{\mathrm{checkpoint}} = \mathcal{K}_{\mathrm{model}}, \tag{106}$$

with exact parameter-shape agreement. Missing keys, unexpected keys, and shape mismatches invalidate export.

## G.2 Static Tensor Interface

The static batch-one interface is

$$\mathbf{X}_t : [1, 6, 400, 400], \tag{107}$$

$$\mathbf{C}_t : [1, 400, 400, 3], \tag{108}$$

$$\mathbf{Y}_t^{\mathrm{cls}} : [1, K_d + 1, 400, 400], \tag{109}$$

$$\mathbf{Y}_t^{\mathrm{reg}} : [1, 6, 400, 400]. \tag{110}$$

Simulation and Real have workload-specific classification-channel counts but share the six-channel double-angle regression interface.

**Table 10: TensorRT FP32 task results for all retained profile–workload engines. Bold marks the selected balanced profile rather than the metric-wise optimum.**

| Workload | Profile | mAP | $AP_{50}$ | $AP_{75}$ | mIoU | Dense mAP | Samples |
|---|---|---|---|---|---|---|---|
| Simulation | Quality | 0.596 | 0.921 | 0.684 | 0.779 | 0.853 | 255 |
| Simulation | Tokens-2 | 0.604 | 0.933 | 0.684 | 0.777 | 0.845 | 255 |
| Simulation | Head-16 | 0.601 | 0.937 | 0.698 | 0.777 | 0.843 | 255 |
| Simulation | **D14** | **0.601** | **0.940** | **0.678** | **0.782** | **0.861** | **255** |
| Simulation | H16-D14 | 0.597 | 0.923 | 0.658 | 0.779 | 0.840 | 255 |
| Real | Quality | 0.106 | 0.394 | 0.017 | 0.533 | 0.460 | 52 |
| Real | Tokens-2 | 0.069 | 0.285 | 0.010 | 0.481 | 0.490 | 52 |
| Real | Head-16 | 0.047 | 0.187 | 0.007 | 0.491 | 0.404 | 52 |
| Real | **D14** | **0.099** | **0.316** | **0.022** | **0.531** | **0.462** | **52** |
| Real | H16-D14 | 0.057 | 0.244 | 0.005 | 0.467 | 0.413 | 52 |

## G.3 Tensor-Level Numerical Regression

For deployed output $\widehat{\mathbf{y}}$ and PyTorch reference $\mathbf{y}$,

$$e_{\max} = \max_i |\widehat{y}_i - y_i|, \tag{111}$$

$$e_{\text{mean}} = \frac{1}{N}\sum_i |\widehat{y}_i - y_i|, \tag{112}$$

$$e_{\text{rmse}} = \sqrt{\frac{1}{N}\sum_i (\widehat{y}_i - y_i)^2}. \tag{113}$$

Spatial classification agreement is

$$A_{\text{argmax}} = \frac{1}{HW}\sum_{u,v} \mathbb{I}\left[\arg\max_c \widehat{Y}^{\text{cls}}_{c,u,v} = \arg\max_c Y^{\text{cls}}_{c,u,v}\right]. \tag{114}$$

Both outputs are checked for NaN and infinity before these statistics are interpreted. Per-tensor error summaries remain artifact-level evidence and are not transcribed manually into this Appendix.

## G.4 Complete-Test Metric Regression

Every sample in the corresponding test split is executed through TensorRT and processed using the same decoder, candidate filter, rotated suppression, and evaluator as its PyTorch reference. For backend $h$, profile $m$, and workload $d$,

$$\Delta \mathbf{q}^h_{m,d} = \mathbf{q}^h_{m,d} - \mathbf{q}^{\text{PyTorch}}_{m,d}. \tag{115}$$

For core metric set

$$\mathcal{M} = \{\text{mAP}, \text{AP}_{50}, \text{AP}_{75}, \text{mIoU}, \text{Dense-mAP}\}, \tag{116}$$

the maximum core deviation is

$$\Delta_{\max}(p,d) = \max_{m\in\mathcal{M}} |m_{\text{TRT}}(p,d) - m_{\text{PyTorch}}(p,d)|. \tag{117}$$

Strict, Metric, Failed, and Pending are qualification categories, not model names. Strict satisfies the registered numerical tolerance. Metric exceeds the strict tolerance but preserves complete-test task metrics within the registered metric-level tolerance. Failed violates qualification, produces non-finite outputs, or cannot complete the test set. Pending lacks complete genuine-input qualification.

# H Complete TensorRT Numerical Regression

## H.1 Simulation Candidate Regression

Table 11 compares the Simulation TensorRT FP32 candidates with their corresponding PyTorch references. All candidates preserve the reference metrics closely; the reported status distinguishes strict numerical regression from task-metric acceptance.

## H.2 Real Candidate Regression

Table 12 summarizes the Real PyTorch-to-TensorRT FP32 regression results. The maximum core delta is the largest absolute difference among mAP, $AP_{50}$, $AP_{75}$, mIoU, and Dense mAP, while yaw is reported separately in degrees.

## H.3 Regression Status

## H.4 FP32 Regression Summary

Table 13 consolidates the numerical qualification outcomes for the eight TensorRT FP32 candidate engines. The two Quality engines serve as workload-specific references and are included in the full-test finiteness audit. All ten FP32 engines complete their respective workload test sets with finite classification and regression outputs. Among the eight candidate comparisons, five satisfy the strict regression criterion and three satisfy the task-metric criterion, with no failed qualification. These results establish the numerical validity of the FP32 deployment artifacts used for the subsequent accuracy and performance comparisons.

## H.5 FP16 and Mixed-Precision Qualification

The unrestricted FP16 engine builds successfully but produces non-finite classification logits and regression outputs on genuine Simulation inputs. The corresponding PyTorch FP32 and TensorRT FP32 executions remain finite on the same tensors, identifying unstable internal FP16 arithmetic as the source of the failure. The unrestricted FP16 engine therefore fails numerical qualification. The constrained mixed-precision engine completes layer audit and short execution tests. Its internal graph still contains substantial FP16 computation, followed by FP32 output reformatting. Full-workload finiteness and task-level regression are required to complete its numerical qualification. Table 14 summarizes the qualification state of the three precision configurations. FP32 is the formal deployment configuration because it completes full-test execution with finite outputs and passes task-level regression.

**Table 11: Simulation PyTorch-to-TensorRT FP32 regression. Differences are computed as TensorRT minus PyTorch from the unrounded metrics and displayed to six decimal places.**

| Profile | ΔmAP | $\Delta AP_{50}$ | $\Delta AP_{75}$ | ΔmIoU | ΔDense | ΔYaw (°) | Status |
|---|---|---|---|---|---|---|---|
| Tokens-2 | −0.000121 | −0.000010 | −0.000061 | −0.000011 | −0.000028 | +0.001082 | PASS_STRICT |
| Head-16 | +0.000605 | −0.000026 | +0.000520 | +0.000083 | −0.000079 | −0.000061 | PASS_STRICT |
| D14 | −0.000100 | +0.000003 | −0.000106 | +0.000102 | −0.000004 | +0.000761 | PASS_STRICT |
| H16-D14 | −0.000944 | −0.000044 | −0.003104 | −0.000046 | −0.000022 | −0.000494 | PASS_METRIC |

**Table 12: Real PyTorch-to-TensorRT FP32 regression.**

| Profile | Max. core \|Δ\| | \|ΔYaw\| (°) | Status |
|---|---|---|---|
| Tokens-2 | 0.002332 | 0.117148 | PASS_METRIC |
| Head-16 | 0.002903 | 0.004214 | PASS_METRIC |
| D14 | 0.000672 | 0.018317 | PASS_STRICT |
| H16-D14 | 0.000472 | 0.001912 | PASS_STRICT |

**Table 13: Final TensorRT FP32 candidate-regression status.**

| Qualification outcome | Count |
|---|---|
| PASS_STRICT | 5 / 8 |
| PASS_METRIC | 3 / 8 |
| FAIL | 0 / 8 |
| Finite candidate outputs | 8 / 8 |
| Finite outputs, all engines | 10 / 10 |

**Table 14: TensorRT numerical qualification by precision configuration.**

| Configuration | Build | Finite | Full test | Qualification |
|---|---|---|---|---|
| FP32 | Yes | Yes | Yes | Formal |
| Unrestricted FP16 | Yes | No | No | Failed |
| Mixed precision | Yes | Pending | No | Pending |

# I Jetson AGX Orin Deployment Protocol

## I.1 Platform Configuration

Table 15 summarizes the fixed platform, execution, and measurement settings used for the formal TensorRT experiments. The same configuration is applied across all retained profiles and both Simulation and Real workloads.

## I.2 Platform Configuration

Table 15 reports the fixed platform and measurement configuration used for all formal TensorRT experiments.

## I.3 Execution Boundaries

FieldRT reports three nested AGX execution boundaries, summarized in Table 16. These boundaries separate engine execution, device-transfer overhead, and the complete point-cloud-to-detection pipeline. The complete-deployment boundary begins with raw point-cloud input and ends when decoded oriented-BEV detections become available. The evaluator-path audit defines a separate diagnostic boundary that additionally includes label loading and target construction. Metric evaluation and external file output remain outside the reported response-time boundaries.

**Table 15: AGX Orin configuration for formal TensorRT evaluation.**

| Item | Configuration |
|---|---|
| Platform | Jetson AGX Orin Developer Kit |
| TensorRT version | 10.3.0 |
| Power mode | MODE_30W |
| Online CPU cores | 8 |
| CPU clock | 1,728 MHz |
| GPU clock | ~612 MHz |
| Active GPU TPCs | 4 |
| EMC clock | 3,199 MHz |
| Batch size | 1 |
| Formal precision | FP32 |
| Warm-up duration | 10 s |
| Measurement duration | 120 s |
| CUDA Graph | Enabled |
| Spin-wait | Enabled |
| Inference streams | 1 |

## I.4 Power and Energy Methodology

Power is reported per named rail. Overlapping rails are not summed. For active power $P_{\text{active}}(t)$ and matched idle baseline $\overline{P}_{\text{idle}}$, idle-adjusted energy per frame is

$$E_{\text{frame}} = \int_{t_0}^{t_1} \max(P_{\text{active}}(t) - \overline{P}_{\text{idle}}, 0)\, dt. \tag{118}$$

For paired saturated runs with mean power and throughput $R$,

$$E_{\text{active/frame}} = \frac{\overline{P}_{\text{active}}}{R}, \qquad E_{\text{dynamic/frame}} = \frac{\overline{P}_{\text{active}} - \overline{P}_{\text{idle}}}{R}. \tag{119}$$

The primary EDP is

$$\text{EDP} = E_{\text{dynamic/frame}} \frac{1}{R}. \tag{120}$$

Component telemetry is interpretive evidence and is not equated with external complete-board DC power.

**Table 16: Operations included in the AGX latency boundaries.**

| Operation | Accelerator only | Transfer inclusive | Complete deployment |
|---|---|---|---|
| Point-cloud loading | | | ✓ |
| ROI filtering | | | ✓ |
| BEV construction | | | ✓ |
| Coordinate construction | | | ✓ |
| Host-to-device transfer | | ✓ | ✓ |
| TensorRT execution | ✓ | ✓ | ✓ |
| Device-to-host transfer | | ✓ | ✓ |
| Output conversion | | ✓ | ✓ |
| Candidate decoding | | | ✓ |
| Rotated suppression | | | ✓ |
| Metric evaluator | Excluded | | |
| External file output | Excluded unless stated | | |

# J Complete AGX Deployment Results

## J.1 Cross-Boundary Profile Summary

Table 17 connects task quality with accelerator-only performance, complete-deployment performance, and idle-adjusted energy across both workloads. D14 provides the strongest cross-workload accuracy–efficiency balance, whereas H16-D14 defines the minimum-latency operating point. Relative to D14, H16-D14 saves only about 1.7 ms in engine median latency while incurring larger losses in strict Simulation localization and Real instance detection. The D14-family engine-level latency reduction also contracts at the complete deployment boundary because preprocessing, transfers, output handling, decoding, and rotated suppression remain substantial, particularly for Real inputs.

## J.2 Engine-Only Performance

Table 18 reports the accelerator-only performance of the retained TensorRT FP32 profiles. The results separate Simulation and Real workloads and include throughput together with median, mean, and tail latency, providing the model-execution reference before data transfers, preprocessing, decoding, and rotated suppression are included.

## J.3 Evaluator-Path Audit

Table 19 reports the evaluator-path timing audit for Simulation and Real. This diagnostic path includes sample fetching, label loading, target construction, TensorRT execution, decoding, and rotated suppression, and therefore complements the accelerator-only measurements in Table 18.

## J.4 Inference-Only Deployment Performance

The inference-only boundary begins with raw point-cloud loading and ends with decoded oriented-BEV detections. It includes BEV and coordinate construction, host-to-device transfer, TensorRT execution, device-to-host transfer, output conversion, softmax, peak extraction, oriented-box decoding, and rotated suppression. Table 20 reports median stage timing and throughput, while Table 21 reports tail latency for the same profile–workload pairs. The D14-family profiles reduce runner latency consistently across both workloads. At the complete-deployment boundary, the improvement is moderated by preprocessing and decoding, which account for a larger share of Real execution. H16-D14 achieves the lowest E2E median latency on both workloads, while D14 retains the stronger cross-workload accuracy–efficiency balance. Tail behavior is stage dependent: H16-D14 gives the lowest Real E2E P95 and P99 despite an isolated increase in its Real runner P99.

## J.5 Power-Window Audit and Rail Statistics

The paired power experiments use 120-second saturated engine-only windows. Table 22 reports the sampling coverage of each run. The VDD_GPU_SOC, VDD_CPU_CV, and VIN_SYS_5V0 rails are analyzed independently throughout the power and energy evaluation. Table 23 reports the matched idle baseline used to derive dynamic power. For each rail $r$, idle-adjusted dynamic power is

$$P_{\text{dyn},r} = P_{\text{active},r} - P_{\text{idle},r}. \tag{121}$$

Table 24 combines the active and idle-adjusted dynamic measurements for all profile–workload pairs. Reporting both values in one table exposes the baseline contribution without duplicating the experiment matrix. Across profiles, GPU/SOC is the dominant measured component rail during engine execution. D14 and H16-D14 reduce energy per frame primarily through higher throughput and shorter execution time rather than through a uniform reduction in instantaneous active power. The independently reported rails provide component-level context for the paired energy and EDP results in Table 25.

## J.6 Energy per Frame and EDP

Table 25 reports the paired VDD_GPU_SOC energy and energy–delay product under saturated, accelerator-only execution. Active energy includes the idle baseline, dynamic energy subtracts the paired idle contribution, and EDP combines dynamic energy with mean engine execution time.

**Table 17: Cross-boundary AGX Orin profile summary. Engine results use the accelerator-only boundary, E2E results use the complete point-cloud-to-detection boundary, and energy is idle-adjusted VDD_GPU_SOC energy per frame. Bold marks the selected balanced profile.**

| Workload | Profile | Eng. qps | Eng. P50 (ms) | E2E fps | E2E P50 (ms) | Energy (J/frame) | TRT mAP |
|---|---|---|---|---|---|---|---|
| Simulation | Quality | 5.755 | 174.582 | 2.540 | 375.324 | 1.124 | 0.596 |
| Simulation | Tokens-2 | 5.945 | 167.328 | 2.557 | 373.406 | 1.123 | 0.604 |
| Simulation | Head-16 | 5.977 | 166.633 | 2.575 | 376.871 | 1.127 | 0.601 |
| Simulation | **D14** | **6.414** | **155.477** | **2.657** | **361.316** | **1.044** | **0.601** |
| Simulation | H16-D14 | 6.445 | 153.777 | 2.653 | 359.488 | 1.037 | 0.597 |
| Real | Quality | 5.717 | 172.215 | 1.752 | 553.412 | 1.145 | 0.106 |
| Real | Tokens-2 | 5.776 | 172.422 | 1.759 | 541.087 | 1.130 | 0.069 |
| Real | Head-16 | 5.851 | 171.609 | 1.759 | 542.914 | 1.118 | 0.047 |
| Real | **D14** | **6.378** | **155.438** | **1.780** | **534.553** | **1.020** | **0.099** |
| Real | H16-D14 | 6.459 | 153.914 | 1.801 | 519.401 | 1.016 | 0.057 |

**Table 18: Complete accelerator-only TensorRT FP32 performance.**

| Workload | Profile | Throughput (qps) | Median (ms) | Mean (ms) | P95 (ms) | P99 (ms) |
|---|---|---|---|---|---|---|
| Simulation | Quality | 5.755 | 174.582 | 173.746 | 179.941 | 180.445 |
| Simulation | Tokens-2 | 5.945 | 167.328 | 168.206 | 173.865 | 175.051 |
| Simulation | Head-16 | 5.977 | 166.633 | 167.303 | 171.846 | 173.628 |
| Simulation | D14 | 6.414 | 155.477 | 155.903 | 160.656 | 161.227 |
| Simulation | H16-D14 | 6.445 | 153.777 | 155.169 | 159.492 | 162.742 |
| Real | Quality | 5.717 | 172.215 | 174.922 | 178.004 | 264.141 |
| Real | Tokens-2 | 5.776 | 172.422 | 173.130 | 173.211 | 256.038 |
| Real | Head-16 | 5.851 | 171.609 | 170.916 | 177.240 | 178.102 |
| Real | D14 | 6.378 | 155.438 | 156.798 | 161.062 | 226.148 |
| Real | H16-D14 | 6.459 | 153.914 | 154.832 | 160.551 | 161.535 |

# K Field-Aware Graph and Traffic Analysis

For profile $p$, the computation is represented as

$$\mathcal{G}_p = (\mathcal{V}_p, \mathcal{E}_p), \tag{122}$$

with tensor-edge annotation

$$\eta_p(e) = (\mathbf{s}_e, q_e, \mathcal{S}_e, \mathcal{D}_e, \mathcal{U}_e, \mathcal{L}_e), \tag{123}$$

where the entries specify shape, bit width, spatial semantics, static or frame dependence, consumer set, and producer-consumer lifetime. For $\mathbf{s}_e = (d_1, \dots, d_K)$,

$$B(e) = \frac{q_e}{8} \prod_{k=1}^{K} d_k. \tag{124}$$

The analysis distinguishes

$$a_e \in \{\text{materialize, retain, stream}\}. \tag{125}$$

External materialization has idealized logical traffic

$$T_{\text{mat}}(e) = [1 + n_{\text{read}}(e)]B(e), \tag{126}$$

whereas full on-chip retention requires

$$M_{\text{retain}}(e) = B(e). \tag{127}$$

Streaming retains only producer-consumer schedule and neighborhood state. In this work, the contract resolution and lifetime-based allocation are applied by hand to each realized kernel.

For edges crossing a device, partition, or explicit materialization boundary,

$$T_{\text{logical}} = \sum_{e \in \mathcal{E}_{\text{boundary}}} \kappa_e B(e), \tag{128}$$

where $\kappa_e$ is the number of logical crossings. Logical edge volume is not physical DDR traffic. Caching, burst alignment, shared consumers, tiling, runtime fusion, fallback execution, and DMA scheduling can change physical traffic. FieldRT therefore distinguishes logical edge volume, DMA-commanded bytes and descriptors, and measured bus or memory-controller traffic.

The Tokens-2 compatibility graph contains 64 ordered regions: 29 DPU, 33 CPU, and two USER-device quantization-entry placeholders. It contains 78 cross-device edges: 41 DPU-to-CPU, 32 CPU-to-DPU, and five involving the USER placeholders. The logical boundary edge-sum upper bound is 261,680,612 bytes per frame, the largest boundary tensor is 15,360,000 bytes, and the median is 1,920,000 bytes. These values show that unsupported-subgraph count omits alternation and boundary volume. They are not measured DDR traffic and are not a causal latency model.

# L FPGA Realization, Validation, and Evidence Boundaries

This appendix presents the programmable-logic realization and validation of FieldRT on the AMD Kria KV260 platform. The hardware study has two complementary roles. First, it characterizes the fragmentation produced when compiled AFN profiles are executed through the standard Vitis-AI compatibility path. Second, it realizes selected geometry-native execution contracts as custom programmable- logic kernels and evaluates their resource feasibility, numerical correctness, and board-level transaction behavior.

The principal custom-PL result is the shared-buffer two-branch multi-dilation denoising kernel. Supporting results cover a single-read Multi-Field Engine (MFE), externally gated streaming fusion, and streaming global reduction. The custom-PL experiments validate the registered arithmetic and dataflow semantics of these operations. Complete AFN task accuracy and end-to-end deployment are reported through the software and TensorRT paths elsewhere in the paper.

**Table 19: Evaluator-path timing audit on Simulation and Real. Stage and end-to-end latencies are reported in milliseconds; throughput is reported in frames per second.**

| Workload | Profile | Fetch P50 | Runner P50 | Decode P50 | E2E P50 | E2E P95 | Throughput |
|---|---|---|---|---|---|---|---|
| Simulation | Quality | 142.028 | 179.360 | 57.173 | 387.515 | 539.729 | 2.479 |
| Simulation | Tokens-2 | 143.041 | 176.668 | 59.336 | 381.849 | 530.149 | 2.511 |
| Simulation | Head-16 | 140.065 | 176.277 | 57.688 | 379.629 | 520.316 | 2.541 |
| Simulation | D14 | 141.206 | 163.302 | 58.512 | 364.161 | 511.251 | 2.627 |
| Simulation | H16-D14 | 143.292 | 163.031 | 57.305 | 367.411 | 519.772 | 2.595 |
| Real | Quality | 279.142 | 182.246 | 73.483 | 558.720 | 780.146 | 1.726 |
| Real | Tokens-2 | 279.212 | 177.105 | 75.552 | 552.697 | 782.252 | 1.735 |
| Real | Head-16 | 276.393 | 176.621 | 91.573 | 558.820 | 791.855 | 1.712 |
| Real | D14 | 274.408 | 164.579 | 88.178 | 534.584 | 770.861 | 1.777 |
| Real | H16-D14 | 278.306 | 163.381 | 83.150 | 540.720 | 785.706 | 1.751 |

**Table 20: Inference-only deployment median timing. Latencies are in milliseconds; throughput is in frames per second.**

| Workload | Profile | Preprocess P50 | Runner P50 | Decode P50 | E2E P50 | FPS |
|---|---|---|---|---|---|---|
| Simulation | Quality | 139.858 | 179.346 | 56.389 | 375.324 | 2.540 |
| Simulation | Tokens-2 | 135.593 | 176.675 | 59.043 | 373.406 | 2.557 |
| Simulation | Head-16 | 134.744 | 176.308 | 56.661 | 376.871 | 2.575 |
| Simulation | D14 | 138.190 | 163.391 | 56.166 | 361.316 | 2.657 |
| Simulation | H16-D14 | 137.153 | 162.994 | 57.471 | 359.488 | 2.653 |
| Real | Quality | 267.063 | 181.950 | 76.759 | 553.412 | 1.752 |
| Real | Tokens-2 | 271.350 | 176.744 | 77.666 | 541.087 | 1.759 |
| Real | Head-16 | 270.193 | 176.104 | 85.254 | 542.914 | 1.759 |
| Real | D14 | 268.752 | 164.014 | 91.829 | 534.553 | 1.780 |
| Real | H16-D14 | 270.533 | 163.127 | 80.049 | 519.401 | 1.801 |

**Table 21: Inference-only deployment tail timing in milliseconds.**

| Workload | Profile | Pre. P95 | Pre. P99 | Run. P95 | Run. P99 | Dec. P95 | Dec. P99 | E2E P95 | E2E P99 |
|---|---|---|---|---|---|---|---|---|---|
| Simulation | Quality | 264.639 | 306.728 | 181.385 | 252.049 | 85.954 | 94.129 | 522.998 | 574.562 |
| Simulation | Tokens-2 | 259.869 | 302.550 | 182.123 | 182.586 | 90.815 | 96.382 | 522.945 | 569.210 |
| Simulation | Head-16 | 260.635 | 300.550 | 181.588 | 183.686 | 82.940 | 87.993 | 513.850 | 562.050 |
| Simulation | D14 | 264.329 | 304.426 | 168.680 | 169.835 | 84.035 | 87.331 | 507.531 | 554.869 |
| Simulation | H16-D14 | 262.769 | 305.911 | 168.278 | 168.921 | 88.354 | 93.878 | 505.143 | 556.257 |
| Real | Quality | 518.515 | 536.774 | 187.638 | 188.853 | 111.497 | 113.881 | 757.566 | 780.922 |
| Real | Tokens-2 | 525.477 | 546.075 | 182.703 | 183.379 | 108.246 | 109.432 | 771.841 | 794.590 |
| Real | Head-16 | 523.289 | 548.220 | 177.104 | 177.198 | 107.157 | 108.864 | 777.045 | 805.707 |
| Real | D14 | 523.468 | 544.971 | 169.457 | 169.778 | 112.071 | 113.005 | 775.675 | 800.572 |
| Real | H16-D14 | 522.510 | 544.514 | 169.441 | 262.889 | 104.363 | 107.598 | 734.214 | 772.120 |

## L.1 Hardware Platform and Evidence Classes

The experiments target the AMD Kria KV260 development platform with device xck26-sfvc784-2LV-c. The device budget used by the hardware analysis is 117,120 LUTs, 234,240 flip-flops, 1,248 DSP slices, 288 $BRAM_{18K}$ blocks, and 64 URAM blocks. Vitis HLS synthesis uses a requested 10-ns clock period unless otherwise stated. HLS-estimated clock periods are not treated as post-route frequencies.

We distinguish four evidence classes throughout the hardware evaluation:

(1) **Compatibility-path evidence** measures execution of compiled Vitis-AI graphs containing DPU and CPU-fallback regions.

**Table 22: Sampling audit for the formal 120-second power windows.**

| Workload | Profile | Samples | Active samples | Coverage (s) | Status |
|---|---|---|---|---|---|
| Simulation | Quality | 1,090 | 1,087 | 120 | PASS |
| Simulation | Tokens-2 | 1,068 | 1,063 | 120 | PASS |
| Simulation | Head-16 | 1,064 | 1,063 | 120 | PASS |
| Simulation | D14 | 1,091 | 1,089 | 120 | PASS |
| Simulation | H16-D14 | 1,090 | 1,088 | 120 | PASS |
| Real | Quality | 1,091 | 1,090 | 120 | PASS |
| Real | Tokens-2 | 1,090 | 1,086 | 120 | PASS |
| Real | Head-16 | 1,090 | 1,089 | 120 | PASS |
| Real | D14 | 1,090 | 1,085 | 120 | PASS |
| Real | H16-D14 | 1,090 | 1,084 | 120 | PASS |

**Table 23: Matched idle-power baseline over 1,103 samples.**

| Rail | Mean idle power (W) |
|---|---|
| VDD_GPU_SOC | 3.936 |
| VDD_CPU_CV | 0.976 |
| VIN_SYS_5V0 | 4.761 |

(2) **HLS synthesis evidence** reports cycles, initiation interval, estimated clock period, and synthesized resource utilization for an individual kernel configuration.
(3) **Board validation evidence** reports an operation executed on the KV260 and compared with an operation-matched software reference.
(4) **Analytical composition evidence** combines independently synthesized kernel reports and is not interpreted as a synthesized or board-measured pipeline.

These evidence classes are kept separate because graph compilation, HLS synthesis, board execution, numerical equivalence, and whole-model deployment answer different architectural questions.

## L.2 Compatibility-Path Fragmentation

The standard Vitis-AI compatibility path partitions both compiled AFN profiles into many alternating accelerator and host regions. Tokens-2 contains 64 leaf subgraphs: 29 DPU regions, 33 CPU-fallback regions, and two USER quantization- entry regions. H16-D14 contains 69 leaf subgraphs: 34 DPU regions and 35 CPU- fallback regions. CPU fallback therefore accounts for 51–52% of the compiled leaf graphs in both profiles.

For Tokens-2, the ordered compatibility graph contains 78 cross-device edges, including 41 DPU-to-CPU and 32 CPU-to-DPU transitions; five additional edges involve USER quantization-entry regions. Shape-based accounting gives a logical boundary edge-sum upper bound of 261,680,612 bytes per frame. The largest individual boundary tensor is 15,360,000 bytes and the median boundary tensor is 1,920,000 bytes.

Logical boundary volume is a graph-derived quantity rather than measured DDR traffic. Caching, shared consumers, burst alignment, runtime fusion, tiling, and fallback scheduling can change the physical traffic. The result is used to expose alternation and materialization pressure that unsupported-operator counts alone do not capture.

## L.3 Profile-Guided Shared-Buffer Denoising

AFN multi-dilation denoising applies several spatial sampling schedules to a common input field. A literal hardware mapping instantiates an independent line buffer for each dilation branch. At production dimensions $C = 64$ and $H = W = 400$, the isolated single-branch HLS estimates require 72, 120, and 216 $\text{BRAM}_{18\text{K}}$ blocks for dilation factors 1, 2, and 4, respectively.

For Tokens-2, whose active dilation set is $\{1, 2, 4\}$, independent buffers therefore require

$$72 + 120 + 216 = 408 \tag{129}$$

$\text{BRAM}_{18\text{K}}$ blocks, or 142% of the 288-block device budget. For the D14 family with active dilations $\{1, 4\}$, independent buffers require

$$72 + 216 = 288 \tag{130}$$

blocks, consuming the complete budget and leaving no capacity for the surrounding design.

FieldRT instead retains the input field once in a line buffer sized for the maximum active dilation. Branch differences are represented by tap schedules and coefficients rather than duplicated activation storage. Both the two- and three-branch production-scale variants consequently use 216 $\text{BRAM}_{18\text{K}}$ blocks, or 75% of the device budget.

The comparison changes the hardware feasibility boundary. The three-branch configuration does not fit under literal buffer replication, whereas the shared-buffer variant preserves 72 $\text{BRAM}_{18\text{K}}$ blocks for the surrounding design. The naive values are sums of isolated per-branch HLS results. The shared two- and three-branch values are production-scale synthesis results. Board validation is complete for the two-branch variant; the three- branch variant remains synthesis validated.

## L.4 Principal Kernel Resource Utilization

The principal board-validated kernel, `denoising_conv_multi2_top`, implements the shared-buffer D1/D4 configuration at $C = 64$ and $H = W = 400$. Table 28 reports its production-scale HLS resource utilization. The design is dominated by line-buffer storage, which occupies 75.0% of the available $BRAM_{18K}$, while its logic and arithmetic resources remain below 8% of device capacity. The HLS-estimated period is 7.458 ns, corresponding to 134.1 MHz. The kernel realizes a checkpoint-derived depthwise-diagonal multi-dilation contract. For each $3 \times 3$ dilation branch, the implementation applies the self-channel coefficients using shared activation storage and an independent accumulation path. Branch fusion, cross-channel mixing, bias, and residual processing remain separate execution objects in the FieldRT spatial architecture. This decomposition isolates the storage and arithmetic behavior of the multi-dilation neighborhood operator, while complete-model quality is measured through the PyTorch and TensorRT evaluation paths.

**Table 24: Mean active and idle-adjusted dynamic power during paired engine-only runs. Each rail is reported independently.**

| Workload | Profile | Active power (W) | | | Dynamic power (W) | | |
|---|---|---|---|---|---|---|---|
| | | GPU/SOC | CPU/CV | 5V0 | GPU/SOC | CPU/CV | 5V0 |
| Simulation | Quality | 10.363 | 1.582 | 8.411 | 6.427 | 0.606 | 3.650 |
| Simulation | Tokens-2 | 10.379 | 1.651 | 8.413 | 6.443 | 0.676 | 3.652 |
| Simulation | Head-16 | 10.618 | 1.551 | 8.489 | 6.682 | 0.575 | 3.728 |
| Simulation | D14 | 10.645 | 1.535 | 8.616 | 6.708 | 0.560 | 3.855 |
| Simulation | H16-D14 | 10.448 | 1.609 | 8.565 | 6.512 | 0.634 | 3.804 |
| Real | Quality | 10.463 | 1.553 | 8.417 | 6.527 | 0.578 | 3.656 |
| Real | Tokens-2 | 10.601 | 1.544 | 8.442 | 6.665 | 0.568 | 3.681 |
| Real | Head-16 | 10.607 | 1.539 | 8.500 | 6.671 | 0.563 | 3.739 |
| Real | D14 | 10.513 | 1.539 | 8.618 | 6.577 | 0.563 | 3.857 |
| Real | H16-D14 | 10.500 | 1.539 | 8.613 | 6.564 | 0.563 | 3.852 |

**Table 25: Paired VDD_GPU_SOC energy per frame and idle-adjusted dynamic energy–delay product under saturated TensorRT FP32 execution.**

| Workload | Profile | Throughput (qps) | Active energy (J/frame) | Idle baseline (J/frame) | Dynamic energy (J/frame) | Dynamic EDP (J s) |
|---|---|---|---|---|---|---|
| Real | Quality | 5.700 | 1.836 | 0.691 | 1.145 | 0.201 |
| Real | Tokens-2 | 5.898 | 1.797 | 0.667 | 1.130 | 0.192 |
| Real | Head-16 | 5.965 | 1.778 | 0.660 | 1.118 | 0.187 |
| Real | D14 | 6.446 | 1.631 | 0.611 | 1.020 | 0.158 |
| Real | H16-D14 | 6.463 | 1.625 | 0.609 | 1.016 | 0.157 |
| Simulation | Quality | 5.717 | 1.812 | 0.688 | 1.124 | 0.197 |
| Simulation | Tokens-2 | 5.735 | 1.810 | 0.686 | 1.123 | 0.196 |
| Simulation | Head-16 | 5.928 | 1.791 | 0.664 | 1.127 | 0.190 |
| Simulation | D14 | 6.423 | 1.657 | 0.613 | 1.044 | 0.163 |
| Simulation | H16-D14 | 6.280 | 1.664 | 0.627 | 1.037 | 0.165 |

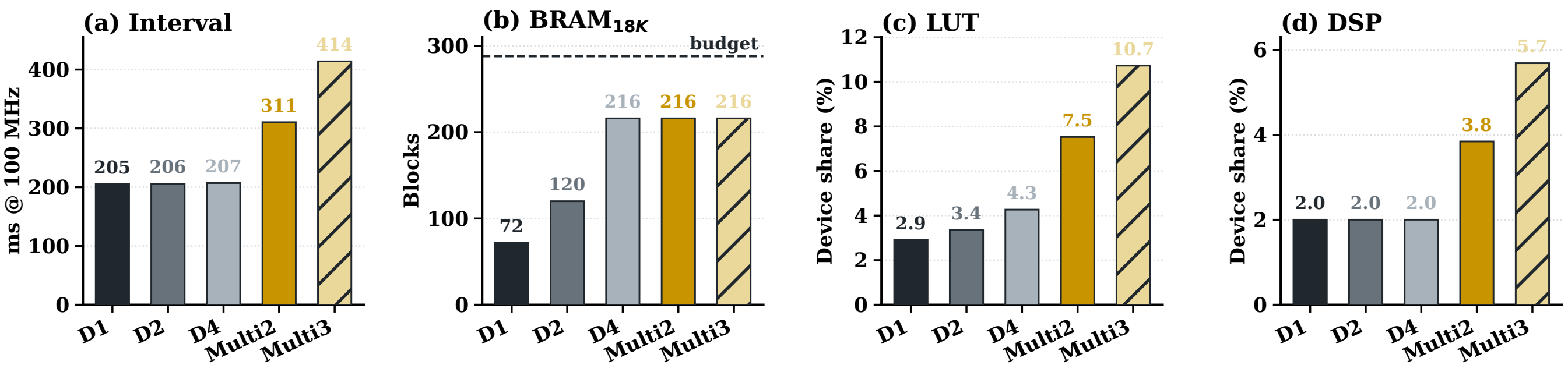

**Figure 14: Production-scale denoising design sweep. The panels compare single-dilation D1, D2, and D4 kernels with the shared-buffer Multi2 and Multi3 variants in initiation interval, $BRAM_{18K}$ usage, LUT utilization, and DSP utilization. Multi2 and Multi3 support additional dilation branches without increasing storage beyond the D4-sized shared buffer.**

**Table 26: KV260 device-resource budget.**

| Resource | LUT | FF | DSP | $BRAM_{18K}$ | URAM |
|---|---|---|---|---|---|
| Available | 117,120 | 234,240 | 1,248 | 288 | 64 |

**Table 27: Compiled compatibility-path partition by leaf-subgraph type.**

| Profile | DPU | CPU fallback | USER | Total |
|---|---|---|---|---|
| Tokens-2 | 29 | 33 | 2 | 64 |
| H16-D14 | 34 | 35 | 0 | 69 |

**Table 28: Production-scale HLS result for the shared-buffer D1/D4 denoising kernel.**

| Resource | Used | Device share |
|---|---|---|
| LUT | 8,814 | 7.5% |
| FF | 3,403 | 1.5% |
| DSP | 48 | 3.8% |
| $BRAM_{18K}$ | 216 | 75.0% |
| URAM | 0 | 0.0% |

## L.5 Production-Scale Board-Level Numerical Validation

The shared-buffer D1/D4 kernel is evaluated using complete checkpoint-derived $64 \times 400 \times 400$ fixtures from both Simulation and Real. Each fixture contains 10,240,000 signed 8-bit input activations, and each branch produces 10,240,000 signed 32-bit outputs, corresponding to 40.96 MB per branch. The same frozen bitstream processes both fixtures with profile-specific activations and coefficients. All four fixture–branch comparisons are bit exact against their operation-matched CPU references. The Simulation and Real D1/D4 outputs each contain 10,240,000 exact matches, with zero mismatches, zero mean absolute error, zero maximum absolute error, and cosine similarity 1.0. These results establish production-scale numerical closure of the implemented shared-buffer multi-dilation contract across both workloads.

## L.6 DMA Topology and Transaction Boundary

The production-scale Real transaction follows

$$\text{AXI DMA}_2 \longrightarrow \text{stream demultiplexer} \longrightarrow \text{multi2 denoising} \longrightarrow \{\text{AXI DMA}_3, \text{AXI DMA}_4\} \tag{131}$$

The two output channels are armed before starting the input transfer. An explicit beat-count register allows the demultiplexer to ignore descriptor- local TLAST events and emit TLAST only for the complete frame.

The input DMA is configured with a 16,383-byte descriptor payload limit. The 10,240,000-byte activation therefore requires 626 descriptors. Each dedicated S2MM output DMA reports the expected 40,960,000 bytes. The final MM2S and S2MM status registers are all `0x00001002`, with no reported wait failure.

**Table 29: Production-scale Real custom-PL transaction.**

| Measurement | Value |
|---|---|
| Input activation | 10,240,000 bytes |
| Input descriptors | 626 |
| Descriptor payload limit | 16,383 bytes |
| D1 output | 40,960,000 bytes |
| D4 output | 40,960,000 bytes |
| Weight loading, before boundary | 0.143 ms |
| Output-DMA setup, diagnostic | 0.007 ms |
| Input-DMA transfer, diagnostic | 429.047 ms |
| Kernel and S2MM completion, diagnostic | 3.052 ms |
| DMA-inclusive accelerator boundary | 441.953 ms |
| Host output dump, outside boundary | 1,574.024 ms |

The 441.953-ms result is an independently measured outer wall-clock boundary. It begins immediately before the two S2MM channels are armed and the input MM2S transfer is started, and ends after both output DMAs report completion. Weight loading occurs before this boundary, and host-side output dumping occurs after it. The segment-level diagnostic intervals are nested observations and do not form an exhaustive additive decomposition of the outer interval.

The measured transaction is input-delivery dominated. The 429.047-ms input-DMA diagnostic interval reflects the current 626-descriptor transfer rather than the arithmetic cost of the denoising operation. Descriptor aggregation, a wider DMA length field, persistent on-chip state, and direct producer-to-kernel streaming are therefore the immediate system optimization opportunities.

## L.7 Supporting Streaming Primitives

*Single-read Multi-Field Engine and streaming fusion.* The MFE reads one feature stream and fans the data into the three geometric field streams consumed by the fusion kernel. At production dimensions $C = 64$ and $H = W = 400$, the MFE–fusion path produces zero mismatches over 10,240,000 elements against a from-scratch CPU reference. The output is SHA-256 identical and reproducible across four of four consecutive board invocations. Production-scale validation currently exercises `GATE_STATIC_SCALAR` with a synthetic uncalibrated gate. The static- channel, sample-scalar, and sample-channel modes are bit exact at tile scale.

*Streaming global reduction.* The reduction primitive is validated with genuine quantized AFN activations. Its CPU reference, HLS C simulation, and KV260 board output agree bit exactly, with zero mismatches over 32 channel outputs, zero mean absolute error, and cosine similarity 1.0. The primitive realizes the streaming global-reduction contract on genuine quantized AFN activations.

## L.8 Analytical Composition Result

Independent HLS reports for MFE with parallelism eight, fusion with parallelism eight, and reduction with channel parallelism eight give maximum intervals of 1,280,005, 1,280,048, and 1,280,019 cycles, respectively. Under ideal streaming composition at 125 MHz, the maximum interval gives an analytical bottleneck estimate of

$$\frac{1{,}280{,}048}{125 \times 10^6} \approx 10.24 \text{ ms.} \tag{132}$$

**Table 30: Summary of validated custom-PL execution contracts. "Production" denotes $C = 64$, $H = W = 400$.**

| Path | Scale | Input provenance | Numerical result | Repetition |
|---|---|---|---|---|
| Shared-buffer denoising D1/D4 | Production | Sim and Real checkpoints | Bit exact per branch | Frozen-bitstream reuse |
| MFE + static-scalar fusion | Production | Synthetic payload/gate | 0/10,240,000 mismatches | 4/4 identical |
| MFE + four gate modes | $C = 8$, $4 \times 4$ | Synthetic tile | 0/128 mismatches per mode | Tile-scale |
| Global reduction | Production dimension | Quantized AFN activation | 0/32 mismatches | Board validated |

**Table 31: Initial KV260 custom-PL power and energy observation.**

| Measurement | Value |
|---|---|
| Transactions | 1 |
| Power monitor | INA260 |
| Measurement rail | Whole-board 5 V |
| Mean idle power | 3.311 W |
| Active-power increment | 0.182 W |
| Incremental energy | 0.365 J |
| Maximum sampling gap | 2.195 s |

The sum of the three independent synthesis reports is 6,430 LUTs, 7,340 flip- flops, 16 DSPs, no BRAM, and no URAM.

This estimate describes an architectural opportunity rather than a measured pipeline. The three kernels have not been jointly synthesized or executed as a single custom-PL chain. The estimate excludes denoising, coordinate modulation, resize, layout conversion, and DPU/PL integration.

## L.9 Initial Custom-PL Power and Energy Observation

The initial custom-PL power characterization uses an INA260 monitor on the KV260 whole-board 5-V rail. Table 31 reports the idle baseline and the incremental power and energy associated with the captured transaction. The measurement captures transaction-level behavior at the whole-board rail. Its artifact record preserves the bitstream, transaction boundary, power samples, and the 2.195-s sampling gap for consistent interpretation of the reported observation.

## L.10 Consolidated Board Measurements

Figure 15 consolidates the measured runtime, transaction diagnostics, payload scale, numerical-validation coverage, repetition evidence, and initial whole-board power observation for the validated custom-PL paths.

## L.11 Implemented Platform and Board-Discovered Hazards

Production-scale board bring-up exposed four dataflow hazards that are not visible in isolated HLS synthesis.

*Shared-ready broadcaster deadlock.* The original broadcaster combined destination readiness while the denoising kernel consumed weights and activation in sequential phases. A destination waiting for a later phase prevented progress in the current phase. The design was replaced by an explicitly controlled stream demultiplexer.

*Route-sampling race.* A free-running demultiplexer could resample its route before software committed the next phase. Route, count, start, done, and idle were moved to a common AXI-Lite control bundle, requiring an explicit software kick per phase.

*Descriptor-local TLAST..* The 14-bit DMA length field forces large tensors into several descriptors. AXI DMA asserts TLAST at each descriptor boundary, but the kernel expects one TLAST for the complete tensor. An explicit beat-count register prevents intermediate TLAST values from terminating the operation.

*Packet-mode output backpressure.* The initial output FIFO exposed packet occupancy only after receiving TLAST, whereas each branch asserted TLAST once per 40.96-MB output. The small FIFO therefore filled before packet completion and backpressured the kernel. The final design uses elastic stream FIFOs and one continuously draining S2MM DMA per branch.

These fixes demonstrate that production-scale correctness depends on the combined kernel, interconnect, packetization, and host-control contract rather than on kernel synthesis alone.

## L.12 FPGA Evaluation Summary

The FPGA experiments establish four main findings. First, conventional compatibility mapping leaves roughly half of the AFN graph in CPU-fallback regions and creates frequent device transitions. Second, profile-guided shared retention changes the three-dilation buffering requirement from an infeasible 408 $BRAM_{18K}$ blocks to a realizable 216-block design. Third, the principal two-branch kernel executes its registered depthwise-diagonal contract bit exactly at production scale for checkpoint-derived Simulation and Real fixtures. Fourth, the current board transaction is limited by descriptor-driven input delivery rather than kernel completion. These results motivate composing the validated contracts into a materialization-avoiding heterogeneous execution path.

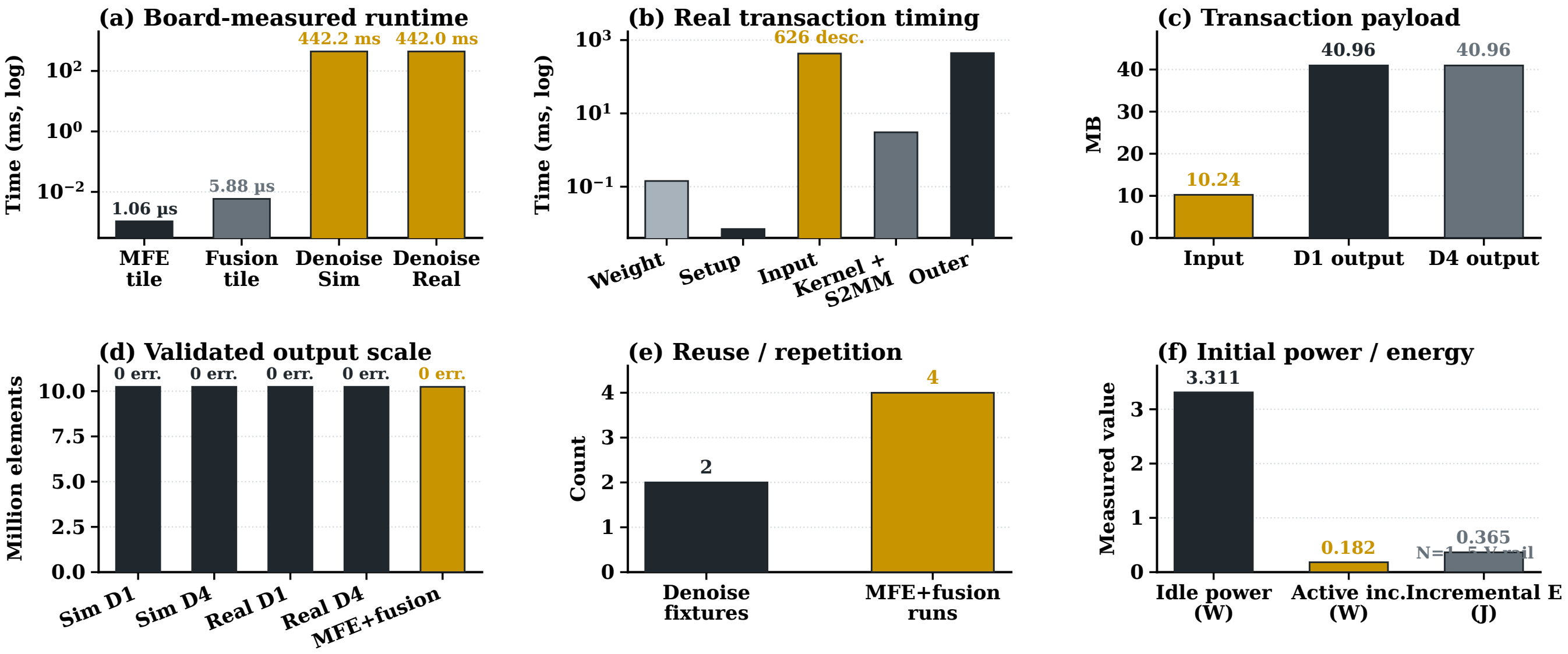

**Figure 15: Consolidated board-level measurements. The panels summarize measured runtime, Real transaction diagnostics, transferred payload sizes, production-scale validation coverage, repeated execution or bitstream reuse, and the initial whole-board power and energy observation. Transaction diagnostics are nested measurements rather than an exhaustive decomposition of the independently measured outer interval.**